\documentclass[fleqn,usenatbib]{mnras}

\usepackage{newtxtext,newtxmath}

\usepackage[T1]{fontenc}

\DeclareRobustCommand{\VAN}[3]{#2}
\let\VANthebibliography\thebibliography
\def\thebibliography{\DeclareRobustCommand{\VAN}[3]{##3}\VANthebibliography}

\usepackage{graphicx}	
\usepackage{amsmath}	
\usepackage[dvipsnames]{xcolor}
\usepackage{xspace}
\usepackage{float}
\usepackage{tikz}
\usepackage{subcaption}
\usetikzlibrary{matrix,chains,positioning,decorations.pathreplacing,arrows}
\usepackage[para]{threeparttable}

\newcommand{\inputs}{\ensuremath{\mathbf{x}}\xspace}
\newcommand{\outputs}{\ensuremath{\mathbf{y}}\xspace}
\newcommand{\preds}{\ensuremath{\mathbf{\hat{y}}}\xspace}
\newcommand{\weights}{\ensuremath{\mathbf{w}}\xspace}
\newcommand{\bias}{\ensuremath{b}\xspace}

\newcommand{\Mass}{\ensuremath{M_{\text{ini}}}\xspace}
\newcommand{\MassError}{\ensuremath{\sigma_{\Mass}}\xspace}

\newcommand{\Zini}{\ensuremath{Z_{\text{ini}}}\xspace}
\newcommand{\ZiniError}{\ensuremath{\sigma_{\Zini}}\xspace}

\newcommand{\Yini}{\ensuremath{Y_{\text{ini}}}\xspace}

\newcommand{\MLT}{\ensuremath{\alpha_{\text{MLT}}}\xspace}

\newcommand{\Age}{\ensuremath{\tau}\xspace}
\newcommand{\AgeError}{\ensuremath{\sigma_{\Age}}\xspace}

\newcommand{\KBa}{\ensuremath{a}\xspace}

\newcommand{\KBb}{\ensuremath{b}\xspace}

\newcommand{\Teff}{\ensuremath{T_{\text{eff}}}\xspace}
\newcommand{\TeffObsError}{\ensuremath{\sigma_{\Teff,\,\text{obs}}}\xspace}
\newcommand{\TeffNNError}{\ensuremath{\sigma_{\Teff,\,\psi}}\xspace}

\newcommand{\Lum}{\ensuremath{L}\xspace}
\newcommand{\LumObsError}{\ensuremath{\sigma_{\Lum,\,\text{obs}}}\xspace}
\newcommand{\LumNNError}{\ensuremath{\sigma_{\Lum,\,\psi}}\xspace}

\newcommand{\FeH}{\ensuremath{\left[\mathrm{Fe}\slash\mathrm{H}\right]}\xspace}
\newcommand{\FeHObsError}{\ensuremath{\sigma_{\FeH,\,\text{obs}}}\xspace}
\newcommand{\FeHNNError}{\ensuremath{\sigma_{\FeH,\,\psi}}\xspace}

\newcommand{\numax}{\ensuremath{\nu_\text{max}}\xspace}
\newcommand{\numaxError}{\ensuremath{\sigma_{\numax}}\xspace}
\newcommand{\dnu}{\ensuremath{\Delta\nu}\xspace}

\newcommand{\muHz}{\,\ensuremath{\mu}\text{Hz}\xspace}
\newcommand{\Modot}{\,\text{M}\ensuremath{_{\odot}}\xspace}
\newcommand{\Lodot}{\,\text{L}\ensuremath{_{\odot}}\xspace}
\newcommand{\dex}{\,\text{dex}\xspace}
\newcommand{\Gyr}{\,\text{Gyr}\xspace}

\newcommand{\Pitchfork}{\textsc{Pitchfork}\xspace}
\newcommand{\Ultranest}{\textsc{UltraNest}\xspace}
\newcommand{\MESA}{\textsc{MESA}\xspace}
\newcommand{\GYRE}{\textsc{GYRE}\xspace}

\newcommand{\kepler}{\textit{Kepler}\xspace}

\usepackage{CJKutf8}
\newcommand{\TLchinesename}{{\begin{CJK}{UTF8}{gbsn}(李坦达)\end{CJK}}}

\title[Asteroseismology of solar-like oscillators]{Asteroseismology of solar-like oscillators: emulating individual mode frequencies with a branching neural network}

\author[O. J. Scutt et al.]{
Owen J. Scutt,$^{1}$ $^{\href{https://orcid.org/0000-0002-7603-3525}{\includegraphics[scale=0.3]{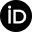}}}$ \thanks{E-mail: oxs235@student.bham.ac.uk (OJS)}
Guy R. Davies,$^{1}$ $^{\href{https://orcid.org/0000-0002-4290-7351}{\includegraphics[scale=0.3]{figs/orcid.png}}}$ 
Amalie Stokholm,$^{1,2}$ $^{\href{https://orcid.org/0000-0002-5496-365X}{\includegraphics[scale=0.3]{figs/orcid.png}}}$ 
Alexander J. Lyttle,$^{3,1}$ $^{\href{https://orcid.org/0000-0001-8355-8082}{\includegraphics[scale=0.3]{figs/orcid.png}}}$ 
\newauthor{
Martin B. Nielsen,$^{1}$ $^{\href{https://orcid.org/0000-0001-9169-2599}{\includegraphics[scale=0.3]{figs/orcid.png}}}$ 
Emily Hatt,$^{1}$ $^{\href{https://orcid.org/0000-0002-1389-1549}{\includegraphics[scale=0.3]{figs/orcid.png}}}$ 
Tanda Li\TLchinesename,$^{4,5,1}$$^{\href{https://orcid.org/0000-0001-6396-2563}{\includegraphics[scale=0.3]{figs/orcid.png}}}$
Mikkel N. Lund,$^{2}$$^{\href{https://orcid.org/0000-0001-9214-5642}{\includegraphics[scale=0.3]{figs/orcid.png}}}$
}
\newauthor{
and Timothy R. Bedding$^{6}$ $^{\href{https://orcid.org/0000-0001-5222-4661}{\includegraphics[scale=0.3]{figs/orcid.png}}}$ 
}
\\
$^{1}$School of Physics \& Astronomy, University of Birmingham, Edgbaston, Birmingham B15 2TT, United Kingdom\\
$^{2}$Stellar Astrophysics Centre, Department of Physics and Astronomy, Aarhus University, Ny Munkegade 120, DK-8000 Aarhus C, DK\\
$^{3}$Advanced Research Computing, University of Birmingham, Edgbaston, Birmingham B15 2TT, United Kingdom\\
$^{4}$Institute for Frontiers in Astronomy and Astrophysics, Beijing Normal University, Beijing 102206, China\\
$^{5}$Department of Astronomy, Beijing Normal University, Beijing, 100875, People's Republic of China\\
$^{6}$Sydney Institute for Astronomy, School of Physics, University of Sydney NSW 2006, Australia\\
}

\date{Accepted XXX. Received YYY; in original form ZZZ}

\pubyear{2026}

\begin{document}
\label{firstpage}
\pagerange{\pageref{firstpage}--\pageref{lastpage}}
\maketitle

\begin{abstract}
Accurately measuring stellar ages and internal structures is challenging, but the inclusion of asteroseismic observables can substantially improve precision.
However, the curse of dimensionality means this comes at a high computational cost when using standard interpolation methods across grids of stellar models.
Furthermore, without a rigorous treatment of random uncertainties in grid-based modelling, it is not possible to address systematic errors in stellar models.
We present \Pitchfork -- a multilayer perceptron neural network with a branching architecture capable of rapid emulation of both classical stellar observables and individual asteroseismic oscillation modes of solar-like oscillators.
\Pitchfork can predict the classical observables \Teff, \Lum, and \FeH with precisions of $5.88$\,K, $0.014\Lodot$, and $0.001\dex$, respectively, and can predict 35 individual radial mode frequencies with a uniform precision of $0.02$ per cent.
\Pitchfork is coupled to a vectorised Bayesian inference pipeline to return well-sampled and fully marginalised posterior distributions.
We validate our rigorous treatment of the random uncertainties -- including the asteroseismic surface effect -- in an extensive hare-and-hounds exercise.
We also demonstrate our ability to infer the stellar properties of benchmark stars -- namely, the Sun and the binary stars 16 Cygni A and B.
This work demonstrates a computationally scalable and statistically robust framework for stellar parameter inference of solar-like oscillators using individual asteroseismic mode frequencies.
This provides a foundation for the treatment of systematics in preparation for the imminent abundance of asteroseismic data from future missions.
\end{abstract}




\section{Introduction}
Characterising distant stars is difficult \citep{Soderblom_2010}.
Estimating stellar fundamental properties -- such as mass, radius, and age -- based solely on photometric, astrometric, or spectroscopic observations poses issues because stellar fundamental properties are poorly constrained by these `classical' observables alone \citep[see e.g.][]{Lebreton_2008, Aguirre_2017, Miglio_2021, Stokholm_2023}.

Asteroseismology -- the study of stellar oscillations -- provides us with a means to improve these constraints.
Including asteroseismic observations can considerably improve fundamental parameter estimation \citep[see e.g. the reviews by][]{Brown_1994, Chaplin_2013, Garcia_2019}.
For instance, combining the classical observables with information from the global asteroseismic measurements -- which describe the overall pattern of oscillations in solar-like stars -- can lead to improvement in mass and radius estimates, and enables age determination with relative precision of 10 to 20 per cent \citep{Chaplin_2014, Aerts_2015}.

In recent years, short-cadence space-based asteroseismic observations from \textit{TESS} \citep{Ricker_2015}, \textit{CoRoT} \citep{Baglin_2006}, and \textit{Kepler} \citep{Borucki_2010} have provided data with enough signal-to-noise to resolve and identify the individual oscillation modes in thousands of solar-like oscillators \citep[see e.g.][]{Hon_2021, Hatt_2023}.
These individual modes of oscillation are more sensitive to the deeper regions of the star than the characteristic oscillation frequency, \numax, and the overtone spacing, \dnu.
Therefore, including them in the inference of stellar fundamental parameters can reduce relative uncertainties on estimates of mass and age by a factor of two or more \citep{Mathur_2012, Aguirre_2017}.

Another issue in estimating stellar fundamental properties is our dependence on models of stellar evolution, which map the stellar fundamental parameters to classical and asteroseismic observables.
Inaccuracies in the model physics assumptions inherent in generating so-called `grids' of stellar models present systematic uncertainties in grid-based inference.
Despite improvements in recent years \citep[e.g.,][]{Rodriguez_2024}, these model grids are still limited in their treatment of mixing and chemical abundances.
Furthermore, improper modelling of the stellar surface produces a significant offset between the modelled and observed oscillation frequencies \citep[the so-called `asteroseismic surface correction';][]{JCD_1988}.
These systematic uncertainties cannot be fully treated until we are confident in our handling of random uncertainties in the estimation of fundamental parameters.

The precision of fundamental parameters estimated via grid-based modelling is not dictated by observational noise alone, but also by the spacing between grid points \citep{Li_2023, Clara_2025}.
This source of error can be reduced by simulating model points `on-the-fly' to match observations using best-fit estimates.
However, this becomes computationally prohibitive if the grid dimensions are increased to include more complex model physics or a large number of individual oscillation modes.
Additionally, the forward modelling dependence of stellar evolution codes means the entire preceding evolutionary track must be calculated at a suitable age resolution to arrive at the target age required to match observations.
Another approach to treat these grid-based random uncertainties is to interpolate precomputed grids of stellar models, which alleviates the forward modelling restrictions of modelling on-the-fly.
Despite promising reported interpolation uncertainties, most interpolation algorithms also become computationally intractable at high dimensions \citep[see e.g. ][]{Rendle_2019, Aguirre_2022}.
This makes it challenging to include individual oscillation modes and varied model physics in the modelling process on a tractable timescale.

Recently, the favourable scaling to higher dimensions of machine learning algorithms has made them more commonplace in the estimation of stellar properties (see e.g.: the random forest regression in \citealt{Bellinger_2016}; the Gaussian process regression in \citealt{Li_Tanda_2022}; and the normalising flows applied by \citealt{Hon_2024} and \citealt{Stone_Martinez_2025}).
In particular, multilayer perceptron neural networks trained as emulators of stellar modelling codes show great promise as an alternative to interpolation.
Neural network emulators can have comparable prediction accuracy to interpolation methods, but are orders of magnitude faster and scale reasonably to higher dimensions (see the comparisons by \citealt{Maltsev_2024,Teng_2025}).
This effective scaling allows consideration of more complex model physics, such as varied mixing \citep{Lyttle_2021} and rotation \citep{Saunders_2024}, as well as the use of individual oscillation modes in making precise age estimates of $\delta$ Scuti oscillators \citep{Scutt_2023}.

In this paper, we present a novel method for modelling solar-like oscillators using Bayesian inference.
We introduce \Pitchfork, a neural network emulator of a grid of models of solar-like oscillators that is capable of rapid predictions of both classical stellar observables and an ensemble of individual modes of oscillation.
We utilise the computational efficiency of \Pitchfork to evaluate the likelihood function in a vectorised Bayesian inference pipeline, which returns posterior samples on the stellar fundamental properties in minutes.
These posteriors are well-sampled, fully marginalised, and demonstrably influenced by the random uncertainties inherent in the stellar modelling process.

In Section \ref{sec:model_grid} we describe the stellar model grid used to train \Pitchfork.
In Section \ref{sec:pitchfork} we briefly introduce the concept of neural networks, and discuss the architecture and prediction precision quantification for \Pitchfork.
In Section \ref{sec:inference} we detail how \Pitchfork is used in a Bayesian inference pipeline and define our priors and likelihood function.
In Section \ref{sec:results} we begin by discussing the objectives of our tests, and then in Section \ref{sec:hare_and_hounds} demonstrate our ability to consistently recover truth values for a population of 250 simulated stars in a hare-and-hounds exercise.
In Section \ref{sec:benchmark_stars} we present results for well-studied benchmark stars -- the Sun and the binary stars 16 Cygni A and B -- and contextualise these against results in the literature while highlighting how this method can be extended in the future.
Finally, in Section \ref{sec:conclusions} we summarise our method and state the main conclusions of our work.

\section{Methods}
\label{sec:methods}
We begin with defining a grid of stellar models that links stellar fundamental parameters to observable quantities.
This grid is required to train our neural network emulator to map between fundamental quantities and observables.
Once trained, we use the emulator to evaluate the likelihood function during Bayesian inference to return estimates of stellar fundamental parameters of an observed star.

\subsection{Grid of Stellar Models}
\label{sec:model_grid}
\begin{figure}
    \centering
    \includegraphics[width=\linewidth]{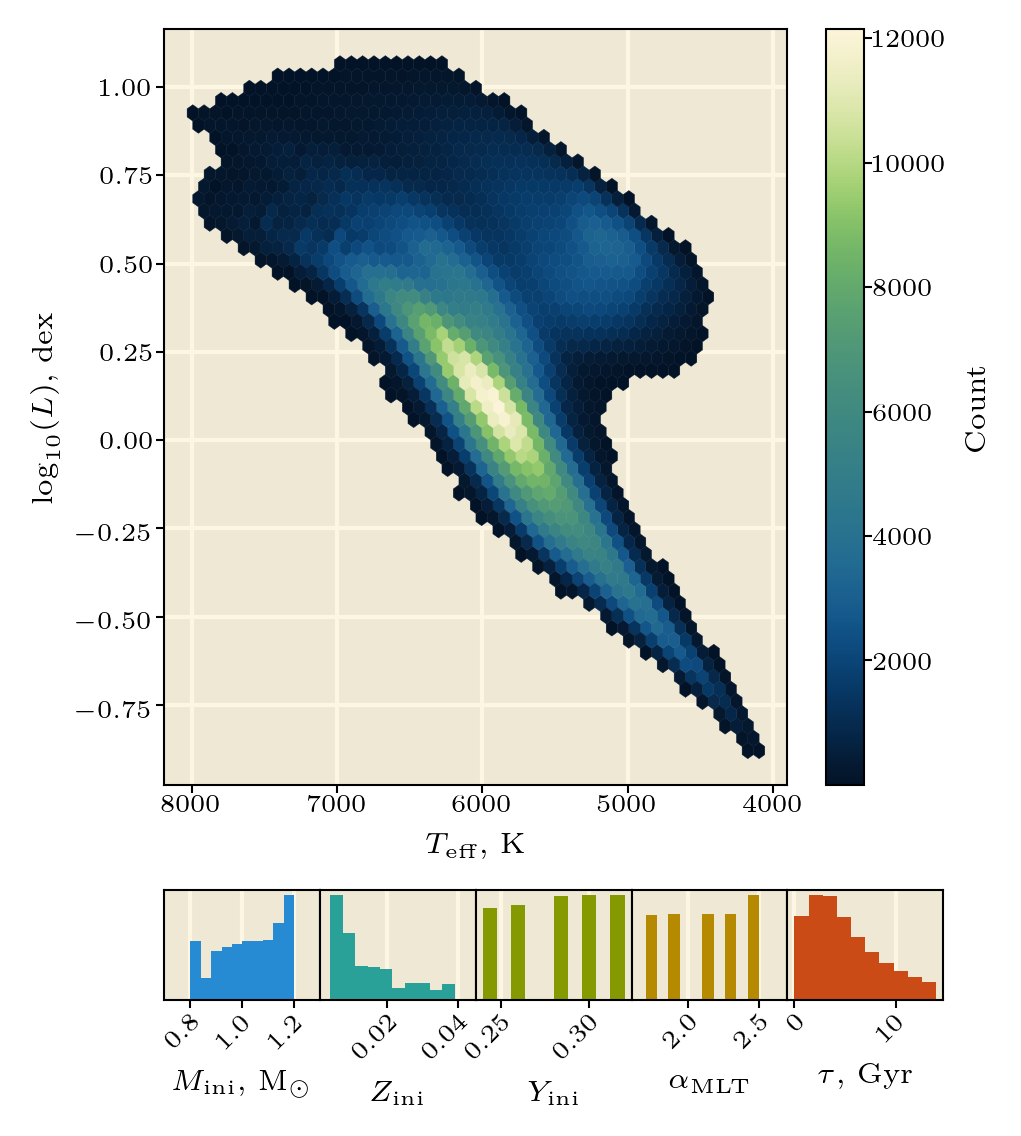}
    \caption{\textbf{Top:} hexbin plot showing counts of model grid points across the HR-diagram. \textbf{Bottom:} distributions of model input parameters used.}
    \label{fig:grid_dists}
\end{figure}

We used the grid of stellar models detailed in \citet{Lyttle_2021} for this work, to which we refer the reader for further details on the chosen model physics. 
Briefly, the stellar model grid was calculated using the \MESA stellar evolution code (version 12115; \citealt{Paxton_2011,Paxton_2013,Paxton_2015,Paxton_2018,Paxton_2019}, \citet{Jermyn_2023}).
The grid considers four model input parameters of which we use 5388 unique combinations: mass \Mass, metallicity \Zini, helium abundance \Yini, and mixing length parameter \MLT.
For each fundamental parameter combination, we evolved forwards in age \Age and sampled along the evolutionary track, resulting in a total of 2448681 stellar models.
Table \ref{tab:grid_details} shows details of the input parameter ranges and step sizes, and Figure \ref{fig:grid_dists} shows the distributions of the \MESA input parameters used.

\begin{table}
	\centering
	\caption{Stellar model grid parameter ranges and step sizes. The step size for \Zini depends on the exact chemical composition of a given model, and the step size in age depends on the rate of change across a track of stellar models -- see \citet{Lyttle_2021} for further details.}
	\begin{tabular}{ccc}
		\hline
		Parameter & Range & Step size\\
            \hline
            \Mass & $0.80 - 1.20\Modot$ & $0.01$ \\
            \Zini & $0.004 - 0.040$ & --- \\
            \Yini & $0.24 - 0.32$ & $0.02$ \\
            \MLT & $1.7 - 2.5$ & $0.2$ \\
            \Age & $0.03 - 14.0\Gyr$ & --- \\
		\hline
	\end{tabular}
    \label{tab:grid_details}
\end{table}

At each step in age, \MESA calculates a series of stellar observables, including the stellar luminosity, \Lum, effective temperature, \Teff, and surface metallicity, \FeH.
These three non-asteroseismic observables are henceforth collectively referred to as the `classical' observables.

The observed power spectrum for a solar-like oscillator shows a series of regularly spaced peaks in frequency, each characterised by a radial order $n$ and angular degree $l$.
To compute the frequency of maximum power, \numax, \textsc{MESA} scales the solar calibrated value with the simulated upper limit in frequency for modes trapped inside the stellar cavity: $\numax \propto g/\sqrt{\Teff}$ \citep{Brown_1991, Kjeldsen_1995}.
In addition, the eigenfrequencies and eigenfunctions of the stellar models were calculated using the \GYRE stellar oscillation code \citep[v5.1;][]{Townsend_2013}.
This provides a host of 35 individual radial oscillation modes (angular degree $l=0$) with radial orders $(6\leq n \leq40)$.
From the individual modes of oscillation, the asteroseismic large frequency separation, \dnu, was calculated by \citet{Lyttle_2021} using the weighted least-squares approach detailed by \citet{White_2011}.

Note that only the individual oscillation modes (collectively referred to as the `asteroseismic' observables in the following text) were used directly in the inference process (see Section \ref{sec:inference}).
The simulated \numax was used for generating realistic observational uncertainty on simulated stars (see Section \ref{sec:hare_and_hounds}), and both \numax and \dnu were used for characterising surface effects (see Section \ref{sec:inference}), meaning neither were used directly as an input for the stellar inference.

\subsection{Scaling and Dimensionality Reduction}
\label{sec:data_prep}
Several steps can be taken before training a neural network to promote faster and more effective training.
For example, scaling all parameters to have a dynamic range close to unity can assist the process of training a neural network emulator \citep{Shanker_1996, Huang_2023}.
We found that the optimal scaling method was taking the base-10 logarithm of all parameters (with the exception of \FeH, which already has units dex) and standardising by subtracting the mean and dividing by the standard deviation.

Reducing the dimensions of the training data before training and re-projecting to the full parameter space within the neural network can also aid the training process \citep[see e.g. ][]{Spurio_Mancini_2022, Scutt_2023, Teng_2025}.
Because the individual mode frequencies have high covariance, and consequently retain the most variance when reduced to fewer dimensions, we performed principal component analysis (PCA) on the asteroseismic observables as follows.
For all models, we calculated the covariance matrix of the individual modes.
The resulting eigenvectors, or `principal components', with largest corresponding eigenvalues explain the majority of the variance of the individual mode frequencies in the model grid.

Replacing the asteroseismic parameters by the reduced dimensions of the principal components presented the neural network with a simpler map from the stellar parameters to the observables.
By training the network to predict these principal components and re-projecting to the full parameter space after prediction, we were able to emulate the entire parameter space with an uncertainty limit determined by the explained variance of the principal components.
We determined how many principal components to include using the explained variance ratio, which describes the percentage of the variance of the observable space present in just the chosen principal components.
We found that including 15 principal components (out of a total of 35) explained all but $6\times10^{-8}$ of the total variance of the individual oscillation modes.
This limit is far lower than even the best predictions made by the emulator (see Section \ref{sec:pitchfork}), and so should not be of concern.

\subsection{\Pitchfork: Neural Network Emulator}
\label{sec:pitchfork}
Grids of stellar models are discretely sampled, which introduces an additional source of systematic uncertainty from interpolating between points.
On the other hand, artificial neural networks are capable of rapid and continuous estimation of the complex functions underlying the dataset on which they are trained.
During training, a neural network is passed a set of these input values from the grid of stellar models, and predicts a set of outputs (the asteroseismic and classical observables).
To emulate the behaviour of \MESA, we trained the network using the inputs: initial mass \Mass, the initial metallicity \Zini, the initial helium abundance \Yini, the mixing length parameter \MLT, and age \Age.
The dynamical range of age for different masses has caused issues for training neural network emulators of stellar evolution code in the past (see e.g. the use of mass-scaled age proxies as inputs in \citet{Lyttle_2021, Scutt_2023}).
However, we found that the neural network architecture we used is capable of predicting to a high precision despite using age as an input.

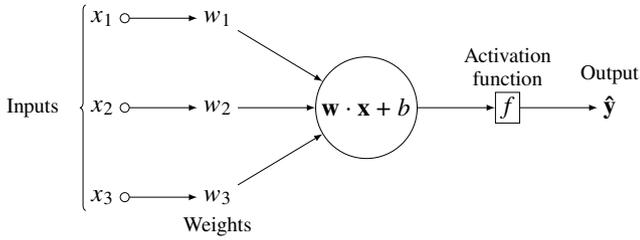
\begin{figure}
    \centering
    \resizebox{\linewidth}{!}{
        \begin{tikzpicture}[
            init/.style={
              draw,
              circle,
              inner sep=2pt,
              font=\Large,
              join = by -latex
            },
            squa/.style={
              draw,
              inner sep=2pt,
              font=\Large,
              join = by -latex
            },
            start chain=2,node distance=13mm
            ]
            \node[on chain=2] 
              (x2) {\Large $x_2$};
            \node[on chain=2,join=by o-latex] 
              {\Large $w_2$};
            \node[on chain=2,init] (sigma) 
              {$\weights \cdot \inputs + \bias$};
            \node[on chain=2,squa,label=above:{\parbox{2cm}{\large \centering Activation \\ function}}]   
              {$f$};
            \node[on chain=2,label=above:{\large Output},join=by -latex] 
              {\Large \preds};
            \begin{scope}[start chain=1]
                \node[on chain=1] at (0,1.5cm) 
                  (x1) {\Large $x_1$};
                \node[on chain=1,join=by o-latex] 
                  (w1) {\Large $w_1$};
            \end{scope}
            \begin{scope}[start chain=3]
                \node[on chain=3] at (0,-1.5cm) 
                  (x3) {\Large $x_3$};
                \node[on chain=3,label=below:{\large Weights},join=by o-latex] 
                  (w3) {\Large $w_3$};
            \end{scope}
            
            \draw[-latex] (w1) -- (sigma);
            \draw[-latex] (w3) -- (sigma);
            
            \draw[decorate,decoration={brace,mirror}] (x1.north west) -- node[left=10pt] {\large Inputs} (x3.south west);
        \end{tikzpicture}
    }
    \caption{The structure of a single neuron. Inputs are combined with weights via a dot product, with a bias term applied. The result is passed through an activation function which scales the neuron output. The neuron output is passed on to the next neuron as an input.}
    \label{fig:neuron_structure}
\end{figure}

For an exhaustive introduction to neural networks, we refer the reader to \citet{Goodfellow_2016}.
In brief, neural networks consist of an input layer, followed by a series of interconnected dense layers that precede a final output layer.
The intermediate layers are populated by individual neurons.
A data point fed into the network during training with a set of inputs, \inputs, will be passed to neurons in the first layer and subject to a linear function of the form
\begin{equation}
    \preds = f(\weights\cdot\inputs + \bias),
    \label{eq:weights_and_biases}
\end{equation}
where \weights is a matrix of weight terms and \bias is a bias term. The result is then passed through some activation function, $f$, before the output, \preds, is passed as an input to all neurons in the following layer.
The structure of a single neuron is shown graphically in Figure~\ref{fig:neuron_structure}.

It is these weights and biases that are tuned during training in order to minimise a defined loss function, which quantifies the magnitude of the residuals between predictions and true values in the training set.
This is repeated for a series of training epochs until the weights and biases are frozen and the network is stored.
For a neural network with a single layer, a single neuron, and a linear activation function (i.e. $\preds = f(\weights \cdot \inputs + \bias) = \weights \cdot \inputs + \bias$), we would be optimising a linear fit between the network inputs and outputs.
By adding many layers each populated with many neurons, and using more complex activation functions, we are able to optimise a generative model for a flexible, highly non-linear function.

By randomly removing a fraction of the entire model grid dataset prior to training, we were able to benchmark a stored network's prediction accuracy on a set of data entirely unseen during training.
We refer to this set-aside data as the `test' set.
To treat overfitting, a common issue in training neural networks in which the network fits to noise in the training set instead of generalising to perform well on data it was not trained on, we also defined a `validation' set which was used as an in-training testing set.
We were able to detect overfitting by monitoring the training and validation loss scores during training -- if the training loss continues to decrease while the validation loss increases or plateaus, we can be confident that the emulator is overfitting to the training set and will not perform well on unseen data.
We found a train/test/validation split of $90/5/5$ per cent of the entire model grid was sufficient, and showed no evidence of overfitting.

\subsubsection{\Pitchfork architecture}
Neural networks are highly customisable.
Examples of this are the number of layers, neurons per layer, and the neuron activation functions, which we collectively refer to as the network \emph{architecture}.
The neural network architecture primarily dictates the maximum flexibility of the network.
An over-complex neural network risks overfitting to the data, and being incapable of translating training success to an unseen test set.
Additionally, computation time during training and prediction will scale rapidly according to network complexity.
Therefore, we seek the simplest possible architecture that still reaches an acceptable level of precision.

Our best-performing neural network, named \Pitchfork hereafter, uses a branching structure to leverage predictive information initially shared between outputs, before splitting and specialising for the classical and asteroseismic observables separately.
We found that the typical architecture, with a linear path from inputs to outputs \citep[such as those used in ][]{Lyttle_2021, Scutt_2023}, was difficult to optimise to promote accurate prediction of both the asteroseismic and classical observables simultaneously.
Furthermore, this allowed us to apply the layer for re-projection from the PCA latent space back to full dimensionality to just the asteroseismic observables.
We used the \textsc{TensorFlow} functional API \citep{Abadi_2015} to construct \Pitchfork.
The other details of the \Pitchfork architecture are given in Table \ref{tab:pitchfork_architecture}.

\begin{table}
	\centering
	\caption{Specifications for our \Pitchfork neural network architecture, designed for this work. The \emph{Stem} values refer to the initial shared fully connected layers, and the \emph{Tine} values refer to the specialised section, treating the classical and asteroseismic properties respectively. The exponential linear unit (ELU) activation function is described in \citet{Clevert_2015}.}
    	\begin{tikzpicture}
	\path
	(0,1) coordinate (A) node[above, inner sep=0]
	{
	\begin{tabular}{cc}
		\hline
		\multicolumn{2}{c}{\textbf{Stem}} \\
		Parameter & Value \\
		\hline
        Input layer units & 5 \\
		Dense layers & 2 \\
		Nodes per layer & 128 \\
		Activation function & ELU \\
		\hline
	\end{tabular}
}
    (-2.1,-0.7) coordinate (B) node[below, inner sep=2] {
	\begin{tabular}{cc}
		\hline
		\multicolumn{2}{c}{\textbf{Tine -- Classical properties}} \\
		Parameter & Value \\
		\hline
		Dense layers & 2 \\
		Nodes per layer & 64 \\
		Activation function & ELU \\
        Output layer units & 3 \\
		\hline
	\end{tabular}
}
(2.1,-0.7) coordinate (C) node[below, inner sep=2] {
	\begin{tabular}{cc}
		\hline
		\multicolumn{2}{c}{\textbf{Tine -- Asteroseismic properties}} \\
		Parameter & Value \\
		\hline
		Dense layers & 6 \\
		Nodes per layer & 128 \\
		Activation function & ELU \\
        Output layer units & 15 \\
        PCA reprojection & $15 \rightarrow 35$ \\
		\hline
	\end{tabular}
};
\coordinate (O) at (0,0);
\coordinate (L) at (-2,0);
\coordinate (R) at (2,0);
  \draw[-{>[scale=2.5]}] (A)--(O) (O)--(L) (L)--(B);
  \draw[-{>[scale=2.5]}] (O)--(R) (R)--(C);
\end{tikzpicture}
	\label{tab:pitchfork_architecture}
\end{table}

\subsubsection{\Pitchfork hyperparameters}
Another example of tunable features in a neural network are the \emph{hyperparameters}, which determine the profile and navigation of the loss landscape.
Examples include The choice of optimiser and corresponding learning rate, the loss function, training batch size, and number of training epochs.
Consideration of neural network hyperparameters is important in training a network -- we might have the perfect architecture to generalise our training set without overfitting, but a poor choice of learning rate would inhibit our ability to ever drop and settle into the global loss minimum.
To find the optimal neural network, we instantiated a grid search routine.
We populated a dense grid with permutations of architectures and hyperparameters and benchmark each with the set aside `test' set to find the best performing network.

\Pitchfork hyperparameters included a Weighted Mean Square Error (WMSE) loss function, defined as
\begin{equation}
    \text{WMSE} = \frac{1}{N}\sum_{i=1}^{N}\left(\frac{y_{i}-\hat{y}_{i}}{\sigma_{i}}\right)^2,
\end{equation}
where $\hat{y}$ is the predicted value output from the final layer, $y$ is the true value, and $\sigma$ is an optional weighting term, summed and averaged over all $N$ output parameters.
We found that using a typical choice for loss function, such as the Mean Squared Error (MSE), resulted in neural networks optimising predictions of just the classical observables.
The WMSE loss function allowed us to set a target level of precision for the neural network by setting the $\sigma$ term to be the desired level of emulator precision on each output.
During training, this greatly incentivised weight and bias tuning, which improved predictions on outputs with uncertainties above $\sigma$.
We typically set these weights to be an order of magnitude lower than estimated observational uncertainties.
The other hyperparameter choices for \Pitchfork are listed in Table \ref{tab:pitchfork_hyperparameters}.

\begin{table}
	\centering
    \begin{threeparttable}
	   \caption{\Pitchfork hyperparameters.}
        \label{tab:pitchfork_hyperparameters}      
        \begin{tabular}{cc}
            \hline
            Hyperparameter & Value \\
                \hline
                Loss Function & WMSE \\
                Optimiser & \textsc{Adam}$^{1}$ \\
                Initial learning rate & $1\times10^{-3}$ \\
                Learning rate decay exponent & $-6\times10^{-5}$ \\
                Minimum learning rate & $1\times10^{-5}$ \\
                Batch size & $2^{15}$ \\
                Epochs & 100000 \\
            \hline
        \end{tabular}

        \begin{tablenotes}
        \textbf{References:} 1 -- \citet{Kingma_2014}
        \end{tablenotes}
        
    \end{threeparttable}
\end{table}

\subsubsection{\Pitchfork evaluation}
\begin{figure}
    \centering
    \includegraphics[width=\linewidth]{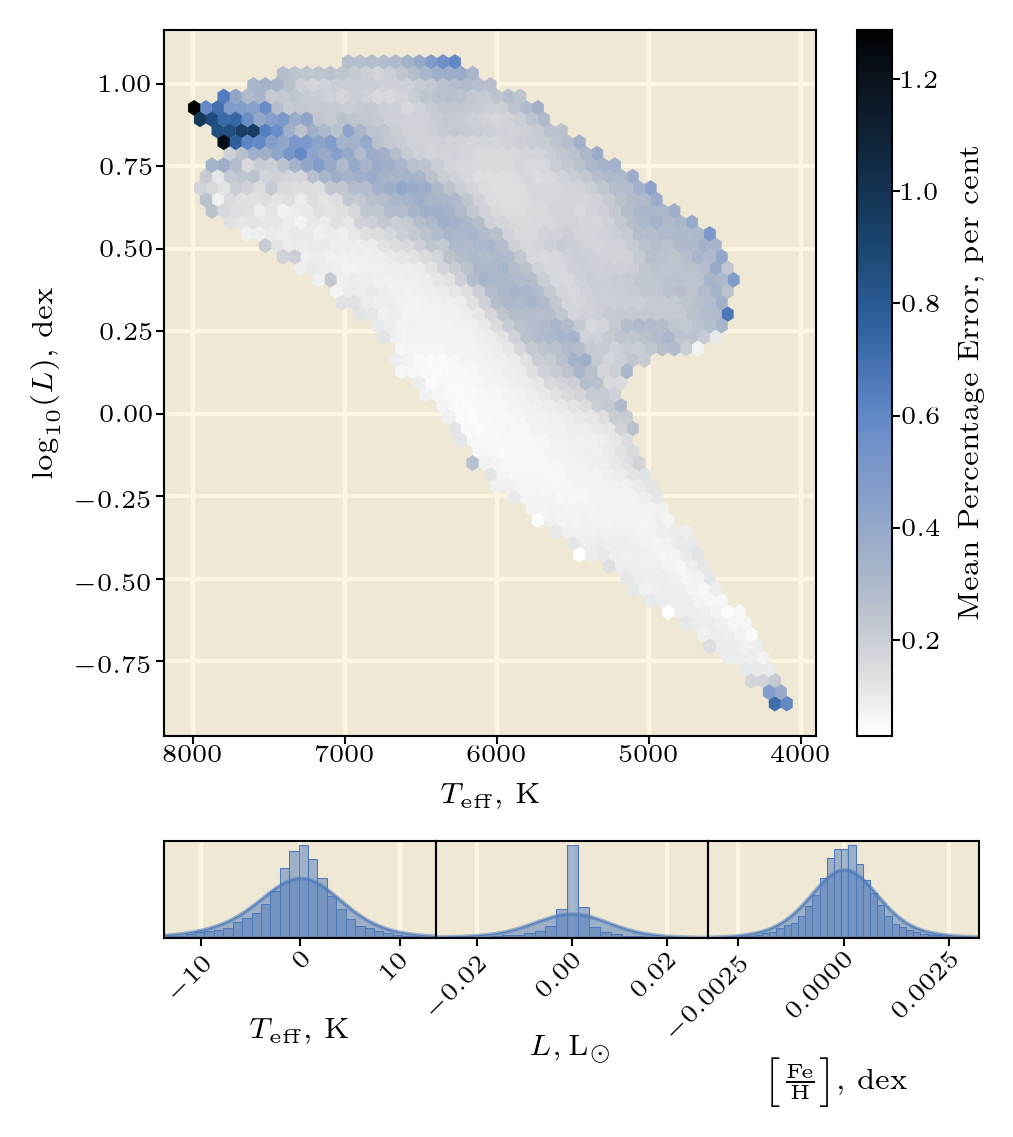}
    \caption{\Pitchfork prediction precision for the classical observables. \textbf{Top:} hexbin plot showing mean percentage error averaged across the classical observables over the HR-diagram. \textbf{ Bottom:} distributions of test set residuals for each classical observable.}
    \label{fig:pitchfork_error_c}
\end{figure}

\begin{figure}
    \centering
    \includegraphics[width=\linewidth]{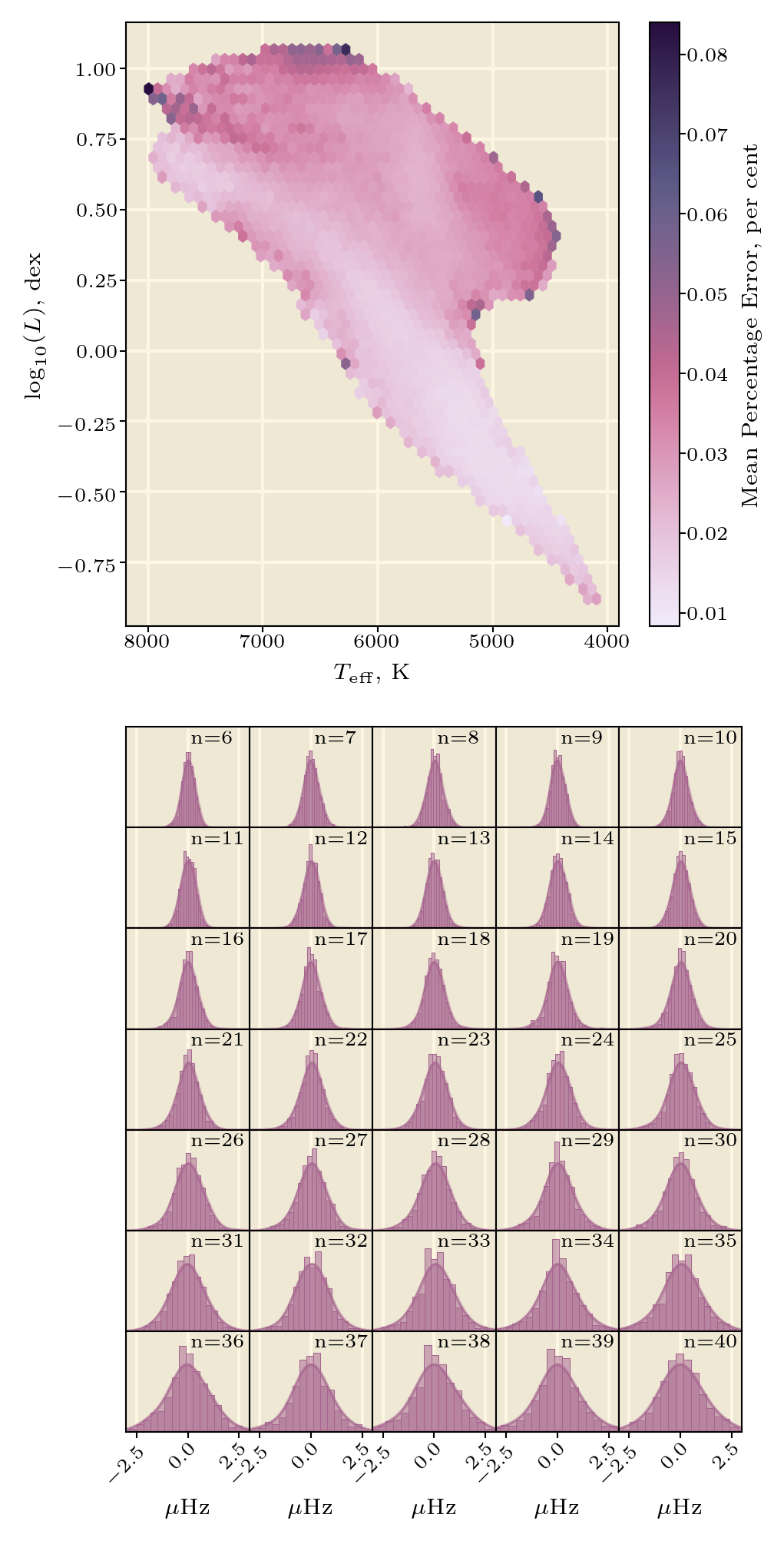}
    \caption{\Pitchfork prediction precision for the individual mode frequencies. \textbf{Top:} hexbin plot showing mean percentage error averaged across all individual mode frequencies (radial orders $(6\leq n\leq40)$) over the HR-diagram. \textbf{Bottom:} distributions of test set residuals on each individual mode frequency, with radial order indicated in the top right.}
    \label{fig:pitchfork_error_a}
\end{figure}

Once trained, we were able to evaluate the success of \Pitchfork on our set aside test set.
For each test point, we called \Pitchfork to predict the outputs and compare to the true value for a set of residuals.
The resulting test set residual distributions provide an estimate for \Pitchfork prediction error over the grid for a given parameter.
The test set residual distributions are shown in Figures \ref{fig:pitchfork_error_c} and \ref{fig:pitchfork_error_a}.

We quote the standard deviation of these distributions as a metric for \Pitchfork prediction uncertainty.
For the classical observables, we report uncertainties of $\TeffNNError = 5.88$\,K, $\LumNNError = 0.014\,\Lodot$, $\FeHNNError = 0.001$\,dex.
The individual mode frequencies have a consistent percentage error on the order of $0.02$ per cent ($\sigma_{n=6,\,\psi} = 0.3\muHz$, $\sigma_{n=40,\,\psi} = 1.1\muHz$).
The full table of \Pitchfork uncertainty across all outputs is summarised in Table \ref{tab:pitchfork_errors}.
We emphasise that these estimates are summary statistics of \Pitchfork performance over the entire grid, and not computed on a model-by-model basis.
In reality, \Pitchfork prediction uncertainty varies across the trained parameter space, as shown in Figures~\ref{fig:pitchfork_error_c} and \ref{fig:pitchfork_error_a}.
When compared to the density of stellar models across the HR-diagram shown in Figure~\ref{fig:grid_dists}, these \Pitchfork residual plots do not show a correlation between regions of higher precision and those of higher training point density.
Instead, we suggest that regions with higher emulator uncertainty are those in which the observables are more sensitive to small changes in the stellar fundamental properties.
As opposed to an interpolator, where we would expect to see diminished precision at the edges of the parameter ranges, \Pitchfork is still capable of predicting to a precision at or exceeding the average precision, even at the edges of the training set ranges.

Considering how the sources of random uncertainty will be accounted for during inference (see Section \ref{sec:inference}), we are aiming for the emulator error to be below the expected observational noise so it is not dominant.
In this regard, the level of emulator precision on the classical observables meets our aims.
On the other hand, \Pitchfork precision on the individual mode frequencies ($\simeq0.5\muHz$) is not insignificant when compared to the levels of observational noise for solar-like oscillators in the trained parameter range, such as for the benchmark stars considered in Section \ref{sec:benchmark_stars}.
This is undesirable, and should certainly be borne in mind when a star has comparable measurement uncertainty on the oscillation modes.
This is a limitation of the method in its current state, and is one that we intend to remedy in the future by utilising the potential for precision increase and point-by-point uncertainty estimation of ensemble deep learning methods \citep[see e.g. ][]{Lakshminarayanan_2017}.

There are examples in the literature of interpolation algorithms for individual mode frequency prediction, or density-scaled proxies thereof, which outperform \Pitchfork, such as in \citet{Rendle_2019} or \citet{Aguirre_2022}.
However, we underline that a comparison to these studies is not like-for-like: the model grid considered in this work is inherently different, and both cases consider variations in fewer dimensions.
A direct comparison would require benchmarking a neural network emulator and an interpolation algorithm over the same grid of stellar models.
For this, we direct the reader to the study by \citet{Maltsev_2024}, who found that their hierarchical nearest-neighbour interpolation algorithm achieves higher predictive accuracy, but that the neural network emulator was two orders of magnitude faster, while still being sufficiently accurate over the parameter space.
We note that the \citet{Maltsev_2024} investigation considered a different stellar model grid and did not attempt emulation or interpolation over individual mode frequencies, and so should not be considered a one-to-one comparison to this study.
Nonetheless, we present our method and results under a similar premise; that the precision reduction on the mode frequencies when using a neural network emulator is easier to remedy than the unfavourable computational scaling of interpolation algorithms.
Furthermore, we demonstrate in the following that this favourable computational scaling renders feasible statistical approaches in which we are confident that the handling of random uncertainties, such as emulation error, is robust.

\Pitchfork took 19 hours to train.
Once trained, it only takes $\sim10\,\text{ms}$ for a single prediction, and is trivial to parallelise, so that \Pitchfork can make $10^6$ predictions in less than $900\,\text{ms}$ on a desktop machine with a GPU\footnote{These timings are for an NVIDIA RTX A4500 GPU. Both the training and prediction times could be reduced considerably by using a high-performance computing cluster with access to GPU(s).}.

This means we have a fast, parallelisable emulator of the \MESA stellar modelling code, free of forward-model dependence, with easily quantified prediction uncertainty which accounts for covariance between outputs.

\subsection{Inference of stellar properties}
\label{sec:inference}

\subsubsection{Priors}
This section details the Bayesian inference pipeline used in this work.
We opted for nested sampling with \Ultranest \citep{Buchner_2021} because nested sampling allows for sampling posterior distributions that are potentially multi-modal or non-Gaussian \citep[for reviews on nested sampling, see ][]{Skilling_2004, Buchner_2023}.
Typically, nested samplers with likelihood functions that are non-trivial to calculate will evaluate the likelihood function sequentially, one sample at a time.
\Ultranest allows vectorised likelihood estimation, which means the likelihood evaluation can accept a large batch of samples simultaneously and return the corresponding likelihoods -- this provides speed gains when the likelihood estimation is parallelisable.
While an interpolator, or modelling on-the-fly, would be difficult to parallelise, neural networks like \Pitchfork are trivial to parallelise.

To infer the values and variances of the fundamental parameters, $\theta$, for a given set of observables of a star, we performed Bayesian inference to sample the fundamental parameter posterior distribution following Bayes theorem:
\begin{equation}
    P(\theta|D) = \frac{\mathcal{P}(\theta) \mathcal{L}(D|\theta)}{\mathcal{E}(D)},
    \label{eq:bayes}
\end{equation}
where $\mathcal{P}(\theta)$ is the prior distribution on the model parameters, $\mathcal{L}(D|\theta)$ is the likelihood of the observed values being returned given the model, and $\mathcal{E}(D)$ is the model evidence which is calculated at each step in the sampling.

The first step was to define $\mathcal{P}(\theta)$, the prior distribution on the stellar fundamental properties.
The functional form of the fundamental parameter prior distributions are shown in Table \ref{tab:prior_funcs}, and we show samples from the prior in appendix Figure \ref{fig:prior_samples}.
These priors were intentionally chosen to be weakly informative, broad, and bounded to the edges of the parameter ranges spanned by the model grid, to avoid sampling outside the emulator's training bounds.
Outside these boundaries, emulation would become extrapolation and our quoted emulator prediction uncertainties would no longer be representative.

\begin{table}
	\centering
	\caption{Prior density functions used. Uniform ($U$) distributions are presented as $U(\textrm{lower limit, upper limit})$. Beta ($\beta$) distributions are given in the form $\beta^{a}_{b}(\textrm{lower limit, upper limit})$, where $a$ and $b$ are the shape parameters of the $\beta$ distribution.} 
    \renewcommand{\arraystretch}{1.2}
	\begin{tabular}{cc}
		\hline
		Parameter & Prior function \\
            \hline
            $\Mass, \Modot$ & $\beta^{5}_{2}(0.8, 1.2)$ \\
            $\Zini$ & $\beta^{2}_{5}(0.004, 0.038)$ \\
            $\Yini$ & $\beta^{2}_{5}(0.24, 0.32)$ \\
            $\MLT$ & $\beta^{1.2}_{1.2}(1.7,2.5)$ \\
            $\Age, \Gyr$ & $\beta^{1.2}_{1.2}(0.03,14)$ \\
            $\KBa, \muHz$ & $U(-10,2)$ \\
            $\KBb$ & $U(4.4,5.25)$ \\
		\hline
	\end{tabular}
    \label{tab:prior_funcs}
\end{table}

\subsubsection{Multivariate Gaussian Likelihood Function}
During nested sampling, samples from the prior are passed as inputs to \Pitchfork, which makes a corresponding prediction.
These predictions are in the observable domain, and can be compared to observed values using a likelihood function.
Typically, the log-likelihood is calculated as a sum of independent normal distributions centred on the observed value, with a width determined by the uncertainties from observational noise and emulator error \citep[see][]{Scutt_2023}.
However, this does not capture any potential covariance between sources of error.
While observational error can be treated as white noise -- fully independent and non-covariate -- other sources of error, such as in predictions from an emulator, can be correlated.

To account for this, we used a multivariate Gaussian likelihood function, which takes the form
\begin{equation}
    \mathcal{L}(\outputs,\boldsymbol{\Sigma}) = (2\pi)^{-\frac{k}{2}} \det(\boldsymbol{\Sigma})^{-\frac{1}{2}} \exp\Bigl(-\frac{1}{2} (\preds-\outputs)^\text{T}\boldsymbol{\Sigma}^{-1}(\preds-\outputs)\Bigl),
    \label{eq:likelihood}
\end{equation}
where \preds is a set of predicted observables for a prior sample, \outputs are the observed values,
$\boldsymbol{\Sigma}$ is the covariance matrix describing the covariance in error (from all sources) of each observed parameter, and $k$ is the rank of $\boldsymbol{\Sigma}$.

\begin{figure}

\begin{subfigure}{.24\textwidth}
  \centering
  \includegraphics[width=\linewidth]{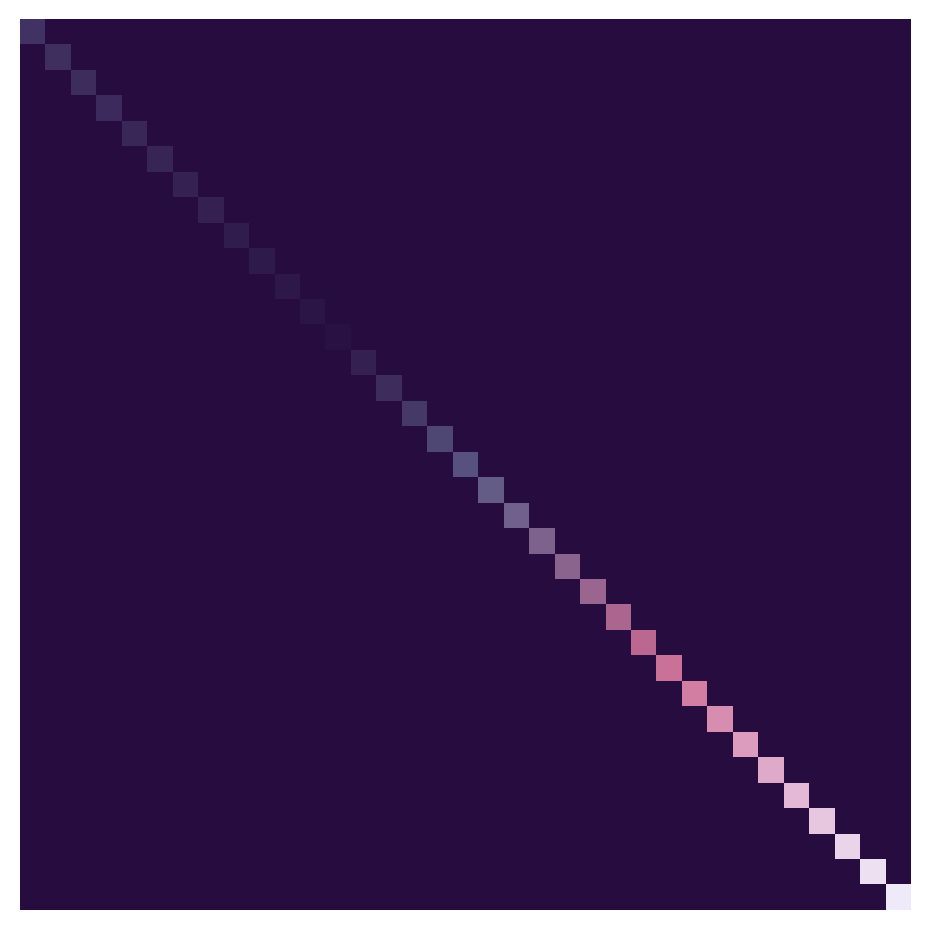}
  \subcaption{$\boldsymbol\Sigma_{\text{obs}}$}
  \label{subfig:obs_cov}
\end{subfigure}%
\begin{subfigure}{.24\textwidth}
  \centering
  \includegraphics[width=\linewidth]{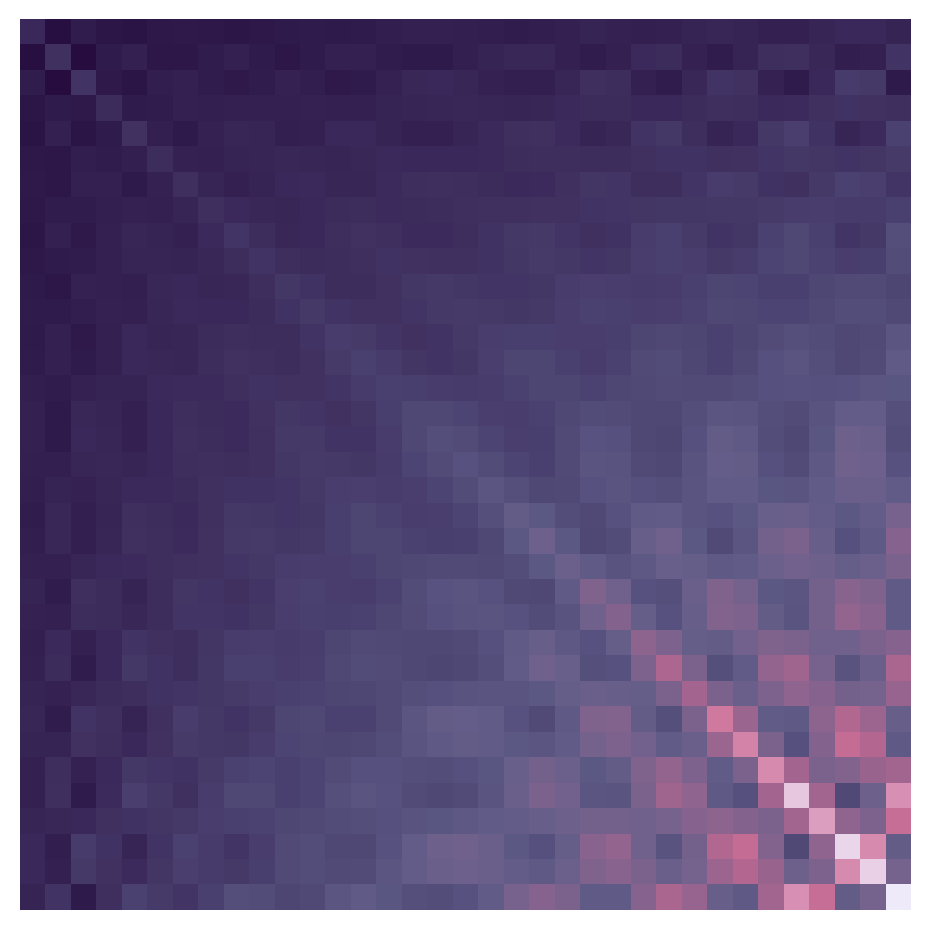}
  \subcaption{$\boldsymbol{\Sigma}_{\psi}$}
  \label{subfig:nn_cov}
\end{subfigure}

\begin{subfigure}{.24\textwidth}
  \centering
  \includegraphics[width=\linewidth]{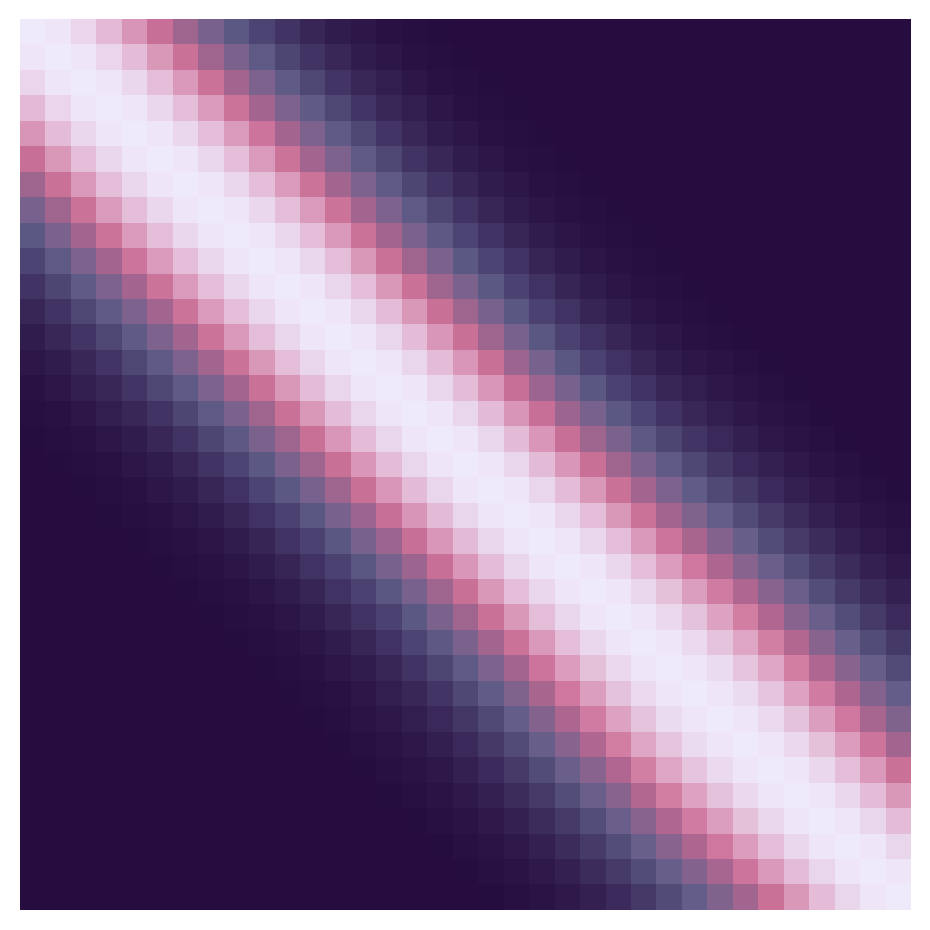}
  \subcaption{$\boldsymbol{\Sigma}_{\text{surf}}$}
  \label{subfig:surf_cov}
\end{subfigure}%
\begin{subfigure}{.24\textwidth}
  \centering
  \includegraphics[width=\linewidth]{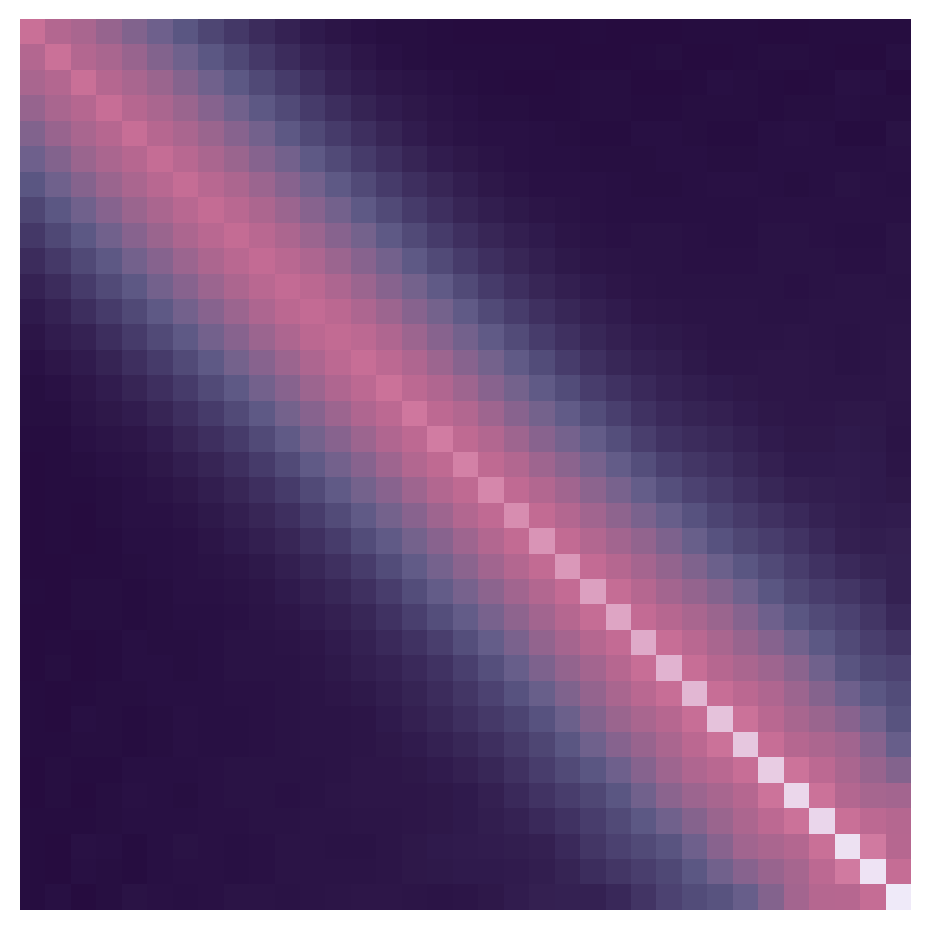}
  \subcaption{$\boldsymbol{\Sigma}$}
  \label{subfig:sigma_cov}
\end{subfigure}

\caption{Examples of the mode frequency component of the different covariances matrices used in defining the multivariate Gaussian likelihood function. \textbf{(a):} the observational noise component $\boldsymbol\Sigma_{\text{obs}}$. \textbf{(b):} the \Pitchfork component, $\boldsymbol{\Sigma}_{\psi}$, from emulation error. \textbf{(c):} the surface correction component $\boldsymbol{\Sigma}_{\text{surf}}$ from the Gaussian process squared exponential kernel. \textbf{(d):} the combined covariance matrix $\boldsymbol{\Sigma}$.}
\label{fig:covariance_matrices}
\end{figure}

The first source of error we consider is the observational uncertainty, which we treat as Gaussian white noise.
For a set of observed quantities, \outputs, with observational uncertainties, $\boldsymbol{\sigma}_{\textrm{obs}}$, the covariance matrix component, $\boldsymbol{\Sigma}_{\textrm{obs}}$, is simply a diagonalised matrix with entries on the leading diagonal equal to the variance ($\boldsymbol{\sigma}_{\textrm{obs}}^2$).
An example for the asteroseismic observables is shown in Figure \ref{subfig:obs_cov}.

The next component, $\boldsymbol{\Sigma}_{\psi}$, treats the error from \Pitchfork.
As detailed in Section \ref{sec:pitchfork}, we determined the grid-wide error of \Pitchfork by calculating the prediction residuals over a set-aside test set.
We then created a covariance matrix for the test set residuals, as shown for the mode frequency predictions in Figure \ref{subfig:nn_cov}.
The leading diagonal of this matrix consists of the variances of the \Pitchfork errors mentioned in Section \ref{sec:pitchfork}, $\boldsymbol{\sigma}_{\psi}^2$.

Note that some \Pitchfork output errors, primarily on the mode frequencies, are highly correlated.
This behaviour is visible in the non-diagonal structure in the covariance matrix on the neural network prediction uncertainties shown in Figure \ref{subfig:nn_cov}.
This behaviour is not a product of the principal component analysis.
Instead, it is a result of the mode frequencies themselves being highly correlated, given they are relatively regularly spaced in the frequency domain (by \dnu).
During training, \Pitchfork is quick to distinguish this regular spacing, but slow to learn how to reproduce the observed deviations from the large frequency separation.

The final component we consider is the error expected from our inability to correctly model the outer layers of a solar-like oscillator \citep{JCD_1988}.
To compensate for this so-called asteroseismic surface effect, multiple corrections of varying complexity can be found in the literature \citep[see e.g. ][]{Kjeldsen_2008, Ball_2014, Li_Y_2023}, each of which models the as offset varying smoothly as a function of mode frequency.
Since \Pitchfork only predicts radial modes (angular degree $\ell=0$) and not, for example, the mode inertias required by the \citet{Ball_2014} prescription, we used the prescription of the surface correction introduced by \citet{Kjeldsen_2008}, which describes the overall frequency shift of the surface effect:
\begin{equation}
    \nu_{n,\,\text{obs}} - \nu_{n,\,\text{model}} = \KBa\left[ \frac{\nu_{n,\,\text{obs}}}{\numax} \right]^{\KBb},
    \label{eqn:surface_correction}
\end{equation}
where $\nu_{n,\,\text{obs}}$ and $\nu_{n,\,\text{model}}$ are observed and modelled radial modes of radial order $n$, respectively.
The denominator is typically a `scaling frequency', for which \citet{Kjeldsen_2008} recommended using the frequency of maximum power \numax.
This surface term prescription models the difference between simulated and observed frequencies as a function of two free variables: \KBa, a multiplicative factor with units \muHz, dictating the magnitude of the frequency shift; and \KBb, an exponent controlling the form of the correction across the frequency spectrum. 

To sample the two surface correction parameters \KBa and \KBb, we used a Gaussian Process (GP) to define a probability distribution on the functional form of the surface correction.
We followed the work of \citet{Li_2023} in using a squared exponential kernel for the GP, which allows for smooth, non-periodic variation in offset as a function of frequency.
We used a constant mean function of zero, as the GP is modelling the correction itself, and not the frequencies with the correction applied.
Our choices for GP kernel length scale and variance determined the profile of the probability distribution and the resulting covariance matrix $\boldsymbol{\Sigma}_{\textrm{surf}}$, which is shown in Figure~\ref{subfig:surf_cov}.

It is worth emphasizing that the flexibility of the GP means we are not solely considering the surface correction parametrised in Equation \ref{eqn:surface_correction}.
When applied in the likelihood function, $\boldsymbol{\Sigma}_{\textrm{surf}}$ defines a probability distribution over \emph{all possible functional forms} of the surface correction.
Our choice of length scale and variance tailor this distribution to prefer \muHz-level deviations that increase smoothly as a function of $n$.
The returned samples for \KBa and \KBb are just the surface correction parameters that are in best agreement with the likelihood according to Equation~\ref{eqn:surface_correction}.

As in \citet{Li_2023}, we used a fixed variance of $4\muHz^2$.  
We determined the length scale on a star-by-star basis by using the returned model evidences to calculate the posterior odds ratio between results obtained using different length scales.
We only considered integer multiples of \dnu as possible length scale values, and note that the flexibility of the GP means it is possible that an incorrect choice for the kernel parameters could potentially lead to the GP absorbing other systematics such as, for example, the helium glitch signature.
Given that we anticipate improvements to the emulation approach that would facilitate a more complex surface correction to be applied (e.g. via emulation of non-radial modes or inertiae), we leave this to future work.
We continue to refer to the frequency-dependent systematics that are being treated by the GP correlated noise model as the `surface term'.
We highlight, however, that there may be other non-surface systematics that are at risk of being absorbed by this approach.

We also compared the evidence when the GP correlated noise model had a variance of zero (i.e. simulating no treatment of correlated noise from imperfect surface correction modelling). 
We found that the data is considerably better explained when using the GP approach than without, and refer the interested reader to Appendix \ref{sec:gp_justification} for more information.

Because $\boldsymbol{\Sigma}_{\textrm{surf}}$ has no bearing on \Teff, \Lum, or \FeH, we padded this array with three corresponding dimensions of zero entries.
Then, since all three error components have identical dimensions, we combined them as follows
\begin{equation}
    \boldsymbol{\Sigma} = \boldsymbol{\Sigma}_{\textrm{obs}} + \boldsymbol{\Sigma}_{\psi} + \boldsymbol{\Sigma}_{\textrm{surf}},
    \label{eqn:sigmas}
\end{equation}
to calculate our combined multivariate Gaussian likelihood covariance matrix, shown in Figure \ref{subfig:sigma_cov}.

With $\boldsymbol{\Sigma}$ calculated, we define our multivariate log-normal likelihood 
\begin{equation}
    \ln\mathcal{L}(\outputs,\boldsymbol{\Sigma}) = \mathcal{K} - \frac{1}{2}\Bigl((\preds - \outputs)^{\text{T}}\boldsymbol{\Sigma}^{-1}(\preds-\outputs)\Bigl),
    \label{eq:log_likelihood}
\end{equation}
where $\mathcal{K}$ is a constant with the form
\begin{equation}
    \mathcal{K} = -\frac{1}{2}\Bigl( \ln(\det(\boldsymbol{\Sigma}))+k\ln(2\pi)\Bigl).
    \label{eq:log_likelihood_constant}
\end{equation}
Given that $\Sigma$ (and $\det\Sigma$) can be pre-calculated, we can define our likelihood constant $\mathcal{K}$ and the inverse $\boldsymbol{\Sigma}^{-1}$ before running the nested sampler, to speed up the likelihood evaluation.
With the prior distribution and likelihood function defined, we can sample the posterior distribution for the fundamental parameters, including the surface correction coefficients \KBa and \KBb, for an observed solar-like oscillator with any number of observed radial mode frequencies from radial orders $6 \leq n \leq 40$.

\section{Results and Discussion}
\label{sec:results}
Here, we present and discuss the results of our work.
In Section \ref{sec:hare_and_hounds} we demonstrate that the method we use is statistically stringent, and that our results are representative of the errors inherent in stellar modelling, by testing on simulated data.
To show that \Pitchfork is capable of emulating the behaviour of real stars, in Section \ref{sec:benchmark_stars} we showcase results for well-studied benchmark stars and compare to literature values.

It is important here to clarify the purpose of this section.
We aim to show that the method we present constitutes a step forward in approaches to stellar modelling, not only in terms of computational tractability, but also through a more robust treatment of random uncertainties and a framework readily extendable to future development.
For one: nested sampling algorithms like \Ultranest offer the ability to capture complex posterior distributions and return the Bayesian model evidence $\mathcal{E}(D)$ explicitly, which is invaluable for model comparison and characterisation of systematics.
These algorithms are typically too computationally intensive to operate over high dimensions when the likelihood evaluation is expensive.
This is not the case when using a trained neural network emulator like \Pitchfork.

The goal of this work was not to present a method that is inherently more accurate or precise than any other.
Indeed, \Pitchfork precision on the mode frequencies is comparable to expected levels of observational noise.
Also, we are bound by the same limitations imposed by our inability to perfectly model stellar evolution as other similar methods.
The point is that the systematic uncertainties inherent in grid-based modelling cannot be addressed until we are confident that the random uncertainties are being handled properly.
Ideally, this would be achieved with a method that is platform-agnostic, adaptable to different grids, and can be extended easily to many dimensions.
We present such a method here.

However, the method in its current form has some limitations: the frequency prediction precision of \Pitchfork is close to expected levels of measurement uncertainty, and we cannot evaluate \Pitchfork precision on a point-by-point basis.
The former is not a limitation specific to neural networks trained as emulators of individual mode frequencies \citep[see the emulator presented by][]{Scutt_2023}.
Rather, it depends on the complexity and relative density of points present in the training set.
One way to alleviate these limitations for this specific grid of stellar models would be to use an ensemble approach.
Furthermore, our method currently only considers radial $(\ell=0)$ oscillation modes.
This limits both the constraint on the fundamental properties as well as prohibiting the use of a more comprehensive prescription of the surface correction used in the GP correlated noise model.
Including non-radial $(\ell\neq0)$ oscillation modes would lift these limitations -- the PCA operation and branching architecture of \Pitchfork should accommodate consideration of non-radial oscillations in the future.

\subsection{Hare-and-Hounds Exercise}
\label{sec:hare_and_hounds}
We begin by demonstrating our ability to recover fully marginalised posterior samples for stellar fundamental parameters of solar-like oscillators by comparing to simulated stars in a `hare-and-hounds' exercise.
A hare-and-hounds exercise is a test in which simulated models (hares) with known fundamental parameters are treated as real stars for the purpose of testing the effectiveness of stellar parameter inference techniques (hounds).
This is widely used in the literature to understand systematics in different modelling approaches \citep[see e.g.][]{Reese_2016, Cunha_2021}.
The goal of this exercise is to validate that the posteriors from the pipeline presented here represent our posterior belief under the assumption that the models are correct.

We discuss results for an exemplar hare in Section \ref{sec:exemplar_hare}, where the returned posteriors were well-constrained and matched the truth values.
In Section \ref{sec:naughty_hare}, we show results for a hare where returned posteriors did not match the truth values, but results were consistent on a population level for different draws of the simulated observational noise applied to the observed parameters used for modelling.
In Section \ref{sec:all_hares}, we show summary statistics for the returned posteriors over all noise draws across a population of 50 hares.

Hares were taken from the set-aside test set, which ensured they had not been seen by \Pitchfork during training.
In order to simulate a realistic observation, we perturbed the observables of the hare as follows: for each observable parameter, we generated a perturbation by sampling from a normal distribution with a mean of 0 and standard deviation of the expected observational uncertainties, for which we used values reflected in the \kepler LEGACY sample \citep{Lund_2017}.
For the classical observables, these were $\TeffObsError = 70\,\text{K}$, $\LumObsError = 0.04\Lodot$, $\FeHObsError = 0.01\,\text{dex}$.

\begin{figure}
    \centering
    \includegraphics[width=\linewidth]{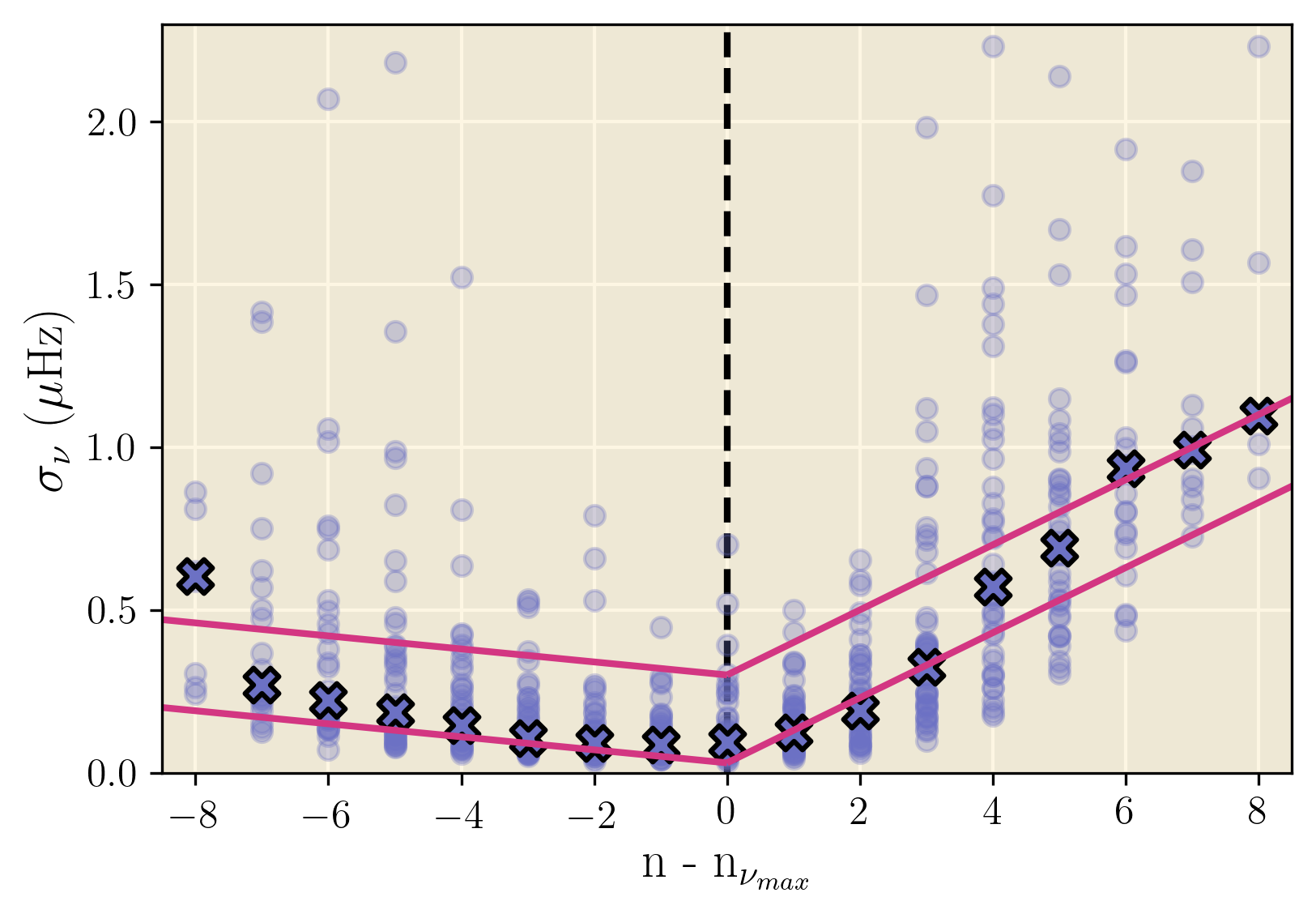}
    \caption{Uncertainties on mode frequencies of the \kepler LEGACY sample \citep{Lund_2017} with inferred \textsc{BASTA} pipeline masses from \citet{Aguirre_2017} below $1.3\Modot$ (purple points) and corresponding medians (purple crosses) shown as a function of $n - n_{\numax}$. The lower and upper pink lines show the minimum and maximum possible frequency uncertainty draws considered in this work. The dotted black line shows $n = n_{\numax}$.}
    \label{fig:legacy_uncs}
\end{figure}

For the asteroseismic observables we took the simulated \numax from the model grid as the mode with highest SNR, and therefore the lowest associated uncertainty.
We defined an uncertainty on \numax, \numaxError, drawn from a uniform distribution between $0.03$ and $0.3\,\muHz$.
Then, for modes observed about \numax we increased the estimated uncertainty on either side as follows
\begin{equation}
    \sigma_{n} = 
    \begin{cases}
        0.1 \times (n-n_{\numax})\muHz, & \text{for }n > n_{\numax} \\
        0.02 \times (n_{\numax}-n)\muHz, & \text{for }n < n_{\numax}
    \end{cases}
    \label{eqn:freq_unc}
\end{equation}
where $\sigma_{n}$ is the uncertainty for the mode frequency of radial order $n$.
Note the increased uncertainty for modes with frequency higher than \numax than those below -- this reflects the decreased mode lifetime (and thus broader peak in the power spectrum) expected for higher frequency modes.
A comparison of these uncertainty draw functions against the frequency uncertainties in the LEGACY sample is shown in Figure~\ref{fig:legacy_uncs}.

Once we had simulated a draw of observational noise, we treated the perturbed values as observed.
We performed inference using the simulated observed values, and compared the recovered posterior with the fundamental parameters used to model the hare.

\subsubsection{Exemplar: Hare 31}
\label{sec:exemplar_hare}
\begin{figure*}
    \centering
    \includegraphics[width=0.7\linewidth]{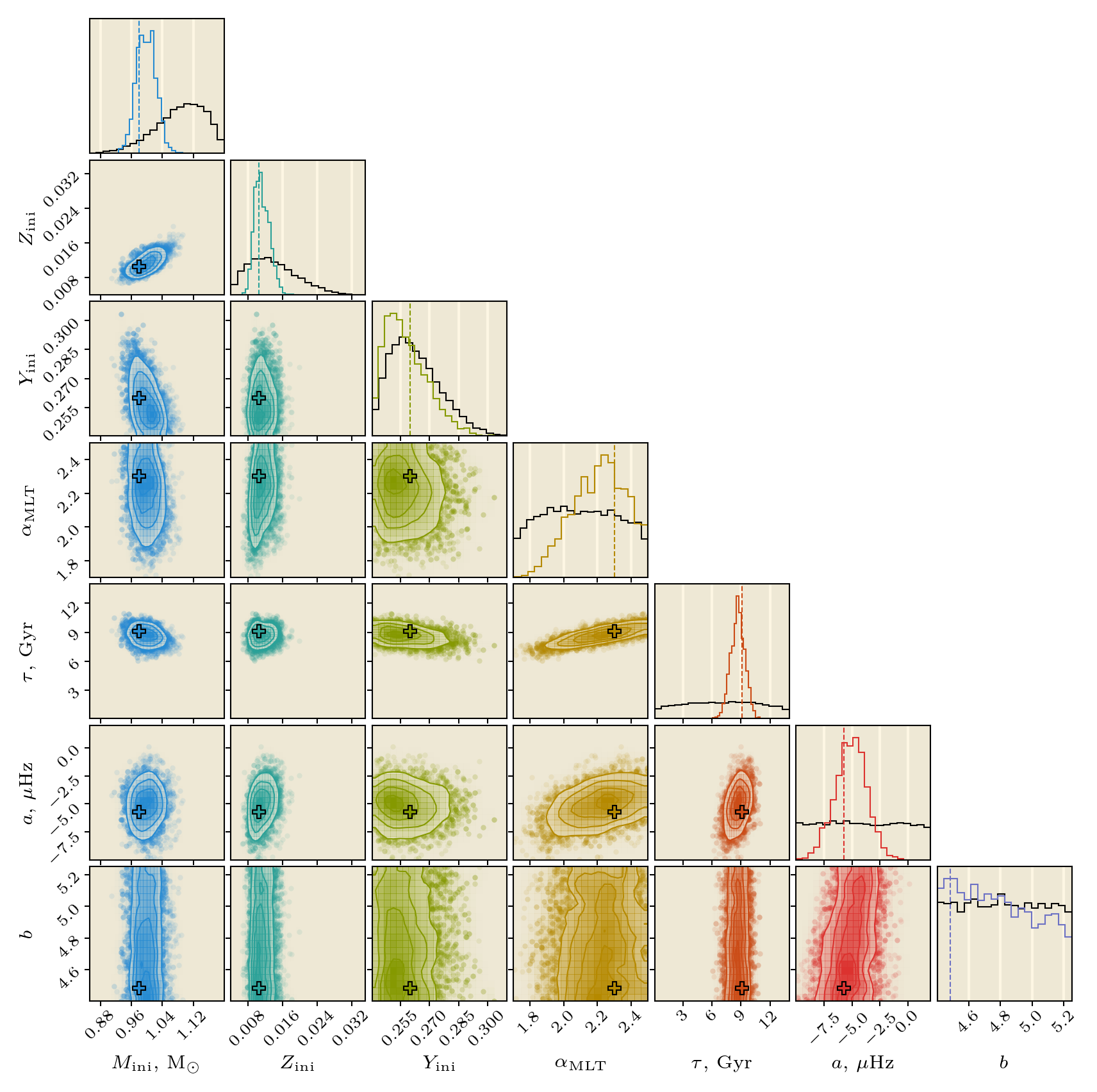}
    \caption{Posterior samples for Exemplar: Hare 31 are shown in colour. The off-diagonal panels show the joint distributions, with a cross plotted to show the truth values used to generate the Hare 31. Marginal distributions are shown in the diagonal panels, with a dotted line showing the truth value, and samples from the prior distribution are shown in black.}
    \label{fig:hare_exemplar}
\end{figure*}

Here we show the results for an exemplar hare: Hare 31 (H31), with posterior samples shown in Figure \ref{fig:hare_exemplar}.
After applying realistic perturbations to the observables, we returned well-constrained marginalised distributions on fundamental parameters with median values ($\Mass = 1.00\Modot$, $\Zini = 0.011$, $\MLT = 2.22$, and $\Age = 8.69\Gyr$) in good agreement with the truth values used to simulate H31 ($\Mass = 0.98\Modot$, $\Zini = 0.010$, $\MLT = 2.3$, $\Age = 9.12\Gyr$) to within 1$\sigma$.

Additionally, these results demonstrate the capability of the GP method for modelling the surface correction. 
We returned posterior samples with a median of $\KBa = -5.09 \pm 1.39\muHz$, which is in agreement with the value used to perturb the simulated mode frequencies ($\KBa = -5.70\muHz$).
The surface term \KBb parameter was poorly constrained for H31.
In fact, \KBb remains prior-dominated for all simulated and real stars we sampled, and so we do not include results for \KBb posterior distributions for the remainder of the paper.
However, \citet{Kjeldsen_2008} demonstrated that the surface term \KBa factor dominates the prescribed correction, while changes to the exponent \KBb term has little effect on the inference of stellar fundamental properties of solar-like oscillators.

We also found that \Yini remains prior-dominated for all results shown here, but we include this in our results because of the presence of covariance in some \Yini joint posterior distributions.
This is to be expected, because accurately constraining \Yini is challenging without characterisation of the asteroseismic glitch signature \citep{Valle_2015, Verma_2019}.

The reader may be concerned that glitch signatures present in the emulated mode frequencies \textemdash prior to correction for surface effects \textemdash could be absorbed by flexibility introduced by the GP correlated noise model, resulting in no meaningful constraints on \Yini. 
However, we found that this is not the case (see Appendix \ref{sec:gp_justification}).

In conclusion, the returned posterior distribution for H31 is well-sampled (5771 samples), fully marginalised, and reflects expected contributions in uncertainty from the emulator, surface correction, and observational noise.
We reported percentage uncertainties on the inferred fundamental parameters of $\MassError = 2.5\,\text{per cent}$, $\ZiniError = 15\,\text{per cent}$, and $\AgeError = 8.5\,\text{per cent}$.

\subsubsection{Effects of simulated observational noise}
\label{sec:naughty_hare}
\begin{figure*}
    \centering
    \includegraphics[width=0.6\linewidth]{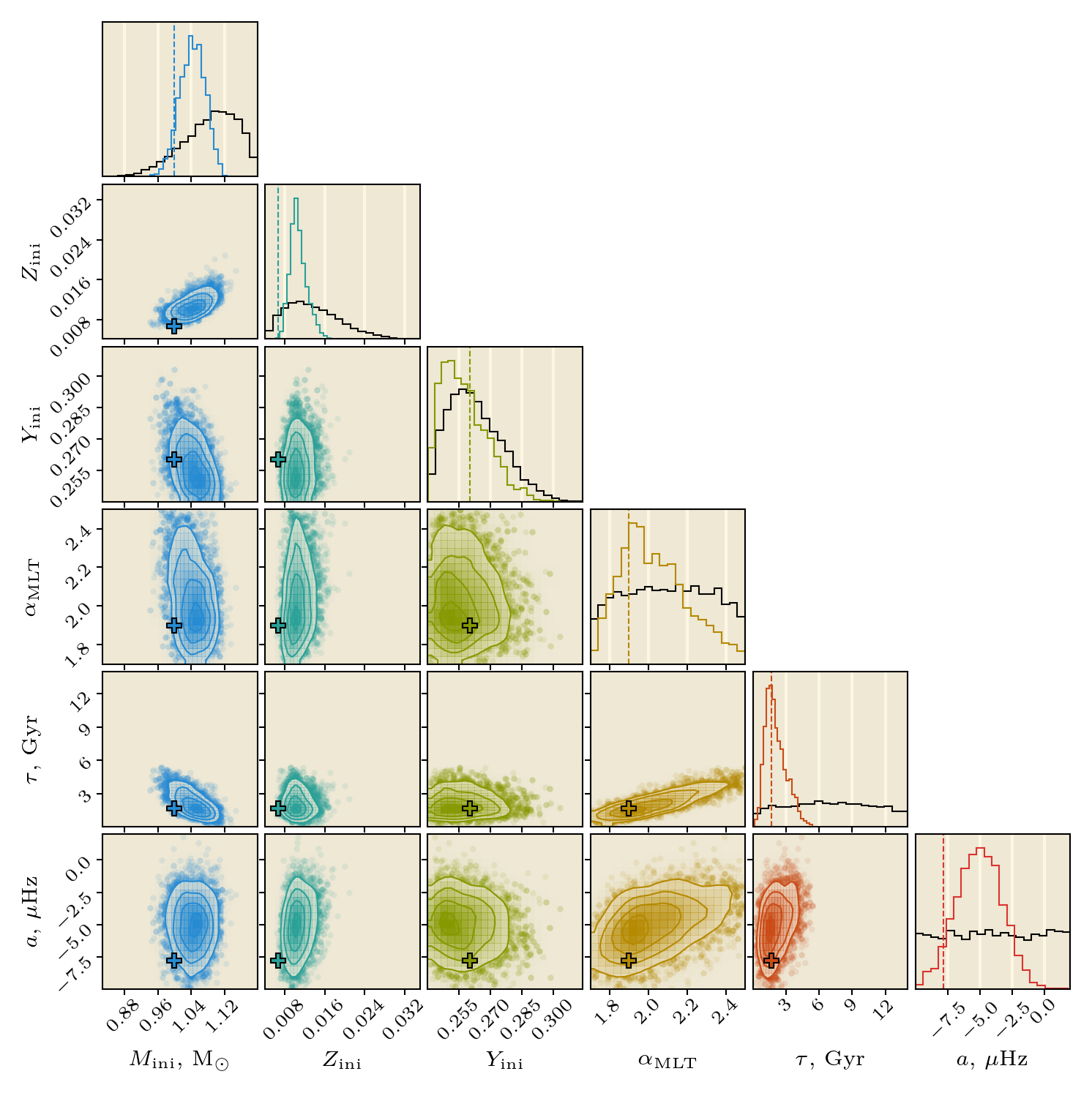}
    \caption{Posterior samples for one draw of simulated observational noise on Hare 43 are shown in colour. The off-diagonal panels show the joint distributions, with a cross plotted to show the truth values used to generate Hare 43. Marginal distributions are shown in the diagonal panels, with a dotted line showing the truth value, and samples from the prior distribution are shown in black.}
    \label{fig:hare_naughty}
\end{figure*}

\begin{figure*}
    \centering
    \includegraphics[width=0.6\linewidth]{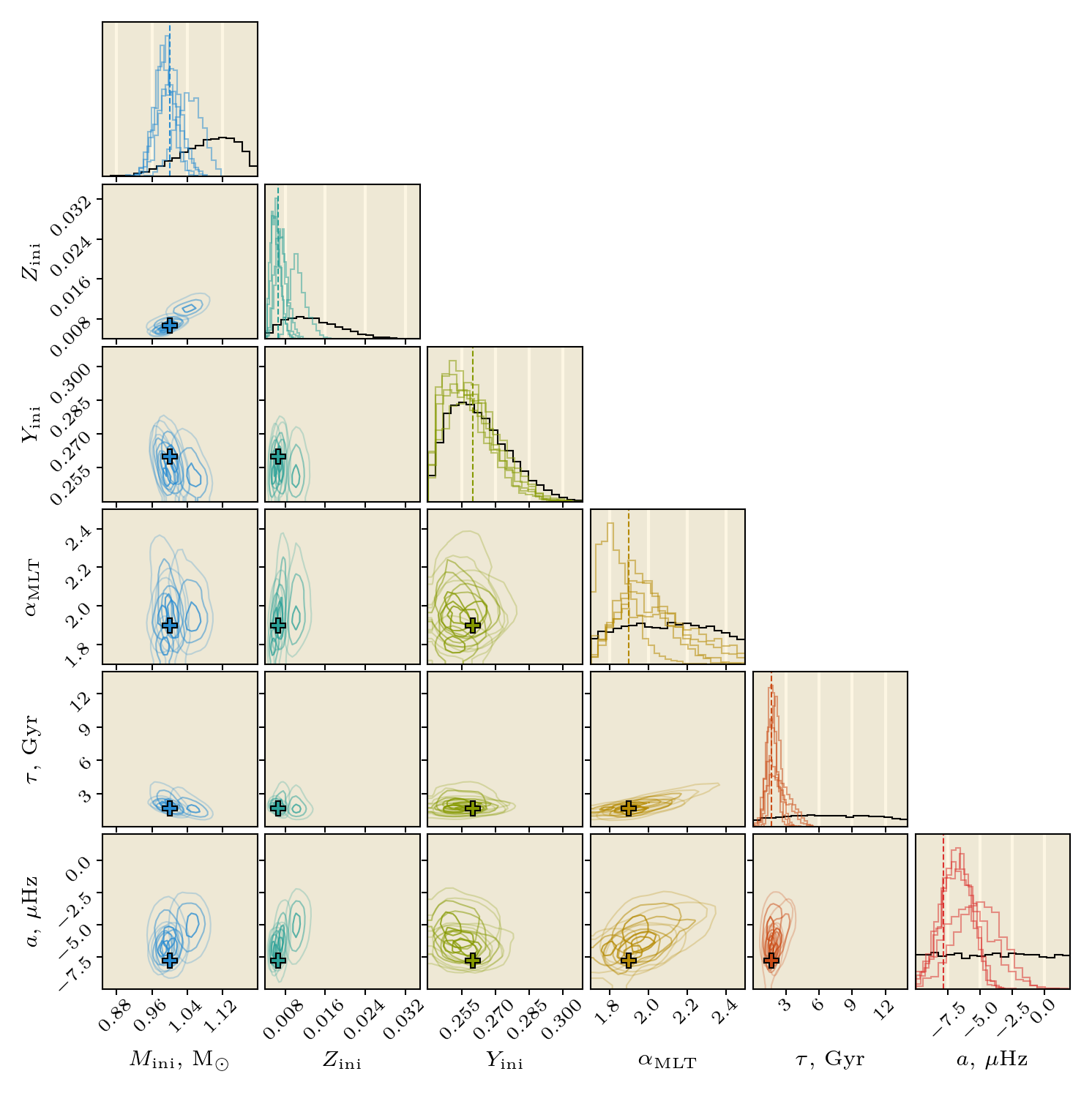}
    \caption{Posterior samples for five draws of simulated observational noise on Hare 43 are shown in colour. The off-diagonal panels show the joint distributions, with a cross plotted to show the truth values used to generate Hare 43. Marginal distributions are shown in the diagonal panels, with a dotted line showing the truth value, and samples from the prior distribution are shown in black.}
    \label{fig:hare_naughty_population}
\end{figure*}

To give confidence in our inferred fundamental parameter values and corresponding uncertainties, we demonstrate our ability to properly treat random uncertainties in our method.
For one realisation of observational noise, the posterior samples for the \Mass, \Zini, and \KBa for Hare 43 (H43) were significantly different from the truth values, as shown in Figure~\ref{fig:hare_naughty}.
This would be of concern if this bias were present in all draws.
However, we also present results for posterior samples of H43 for a further 4 different realisations of observational noise in Figure \ref{fig:hare_naughty_population}.
The remainder of these noise realisations returned posterior samples that were consistent with the truth values used to simulate H43.

The draw of the observational noise used to sample the posterior in Figure \ref{fig:hare_naughty} is responsible for the skewed posterior samples -- the discrepant realisation reduced \Teff by $190\,\text{K} ($$2.7\times\TeffObsError$) and increased \FeH by $0.17\,\text{dex}$ ($1.5\times\FeHObsError$).
Proper treatment of the random uncertainties in stellar modelling should result in a measurable effect on the inferred posterior distributions, and we demonstrate that here.

\subsubsection{All Hares}
\label{sec:all_hares}
To showcase our method on a wider level, we also show results for samples from a population of 50 hares, each with 5 different realisations of observational noise.
For each marginalised posterior distribution, we calculated the posterior $z$-score by subtracting the posterior mean from the `true' value used to simulate the hare and dividing by the posterior standard deviation.
This posterior $z$-score is an indicator of the success of our inference method in our hare-and-hounds exercise at a population level: if our treatment of emulator uncertainty were perfect, and the returned marginalised posterior distributions were all Gaussian, the $z$-score distribution would be consistent with a normal distribution with zero mean and unit standard deviation.

\begin{figure*}
    \centering
    \includegraphics[width=\linewidth]{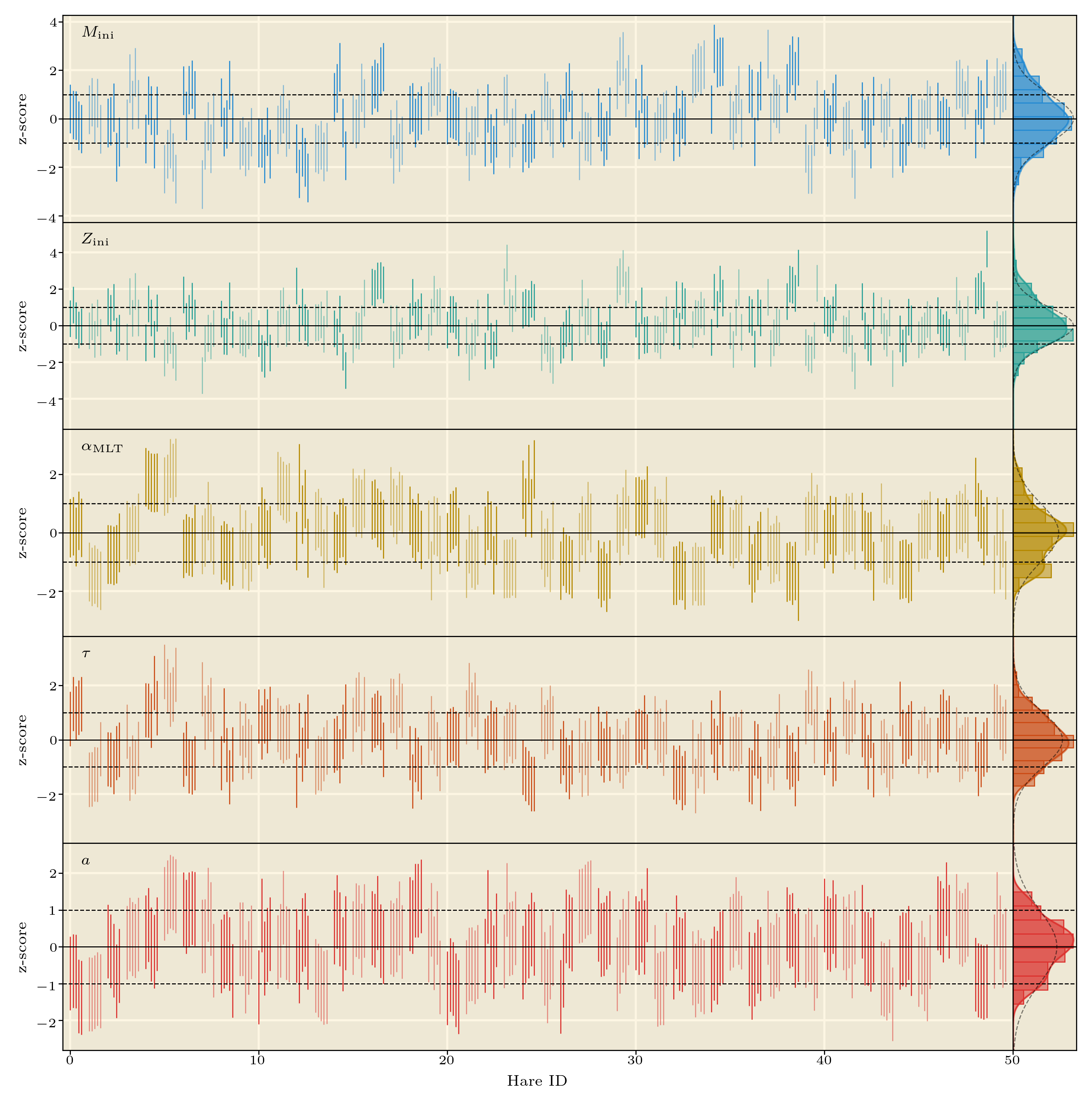}
    \caption{Posterior $z$-scores over a population of hares, each with five draws of observational noise. Each vertical line represents one draw of the observational noise. We alternate the saturation of $z$-score lines for clarity. The histograms and kernel density estimates of all returned $z$-scores are shown on the right hand side, with the target $\mathcal{N}(0,1)$ shown by the dotted grey line.}
    \label{fig:hares_population}
\end{figure*}

In Figure \ref{fig:hares_population}, we show the 1$\sigma$ spread of some of the sampled parameter $z$-scores and the distributions of the $z$-score means across the population.
On this large population level, our returned $z$-scores for \Mass, \Zini, \MLT, \Age, and \KBa are consistent with an $\mathcal{N}(0,1)$ distribution, as can be seen in the histograms in Figure \ref{fig:hares_population}.
We do not show results here for \Yini or \KBb because these parameters are rarely well enough constrained to return posterior distributions with a close-to-Gaussian profile.

We note a trend for self-consistency among the $z$-scores for different draws of the observational noise for a given hare.
This effect is present for two reasons.
One is the inherent assumption that the posteriors are close to Gaussian when calculating the $z$-score.
In reality, this assumption breaks down when we are sampling posteriors that are centred in a region of parameter space that is close to the edge of the prior distribution; the returned posterior may peak at a value centred on the truth, but the mean of the distribution will be skewed towards the centre of the prior. 
This could be solved by only drawing hares that lie comfortably within our prior distribution, but this would neglect to test our method in the entire prior space.
Alternatively, we could define priors that exceed the bounds of the grid on which the emulator was trained.
This would allow the emulator to extrapolate on out-of-distribution data at an inflated uncertainty, which would be poorly represented by our projected emulation uncertainty covariance matrix used in the likelihood function.

The second contributing factor is due to biases in the neural network's predictions.
For example, if the emulator is prone to bias in effective temperature prediction in a region of fundamental parameter space, then this will be reflected in the exploration of the likelihood function during nested sampling. 
We tested for this contribution by using emulated observables for a set of fundamental parameters as inputs for the inference pipeline instead of using simulated values.
A population of emulated hares, which we call `emus', was used for a population-level test of our inference pipeline much like the hare-and-hounds exercise above.
This emu-and-hounds exercise thus measured the bias inherent in the emulator predictions, while the traditional hare-and-hounds exercise tested for bias related to the injected observational uncertainty.
As seen in Figure \ref{fig:10_zscores} we find that the emulator bias is not the dominating factor -- the posterior $z$-scores for the first 10 hares and emus look close to identical.
Regardless, this contribution should be addressed, and we aim to do so in future work by using an ensemble of emulators in place of the single emulator used in this work.
By using ensemble methods, we can take the mean prediction of the ensemble for a given point as the prediction, and the error on the mean as a point-by-point uncertainty metric.
For a large ensemble, the individual emulator biases cancel out and the ensemble prediction should be free of systematic bias across a region of parameter space.

\subsection{Application to Benchmark Stars}
\label{sec:benchmark_stars}
Having tested our method on a set of simulated stars, here we present results for three real stars and contextualise our results against literature values.
Section \ref{sec:sun} shows posterior samples for the Sun and showcases the diagnostic potential of a posterior predictive check.
Section \ref{sec:16cyg} shows results for the binary system 16 Cygni A and B, treated as individual stars, to demonstrate recovery of the consistent \Zini and \Age expected from a binary system.

\subsubsection{The Sun}
\label{sec:sun}
Here we present our results for the Sun. 
Table \ref{tab:solar_classicals} shows the adopted classical observables used, and Table~\ref{tab:solar_asteroseismic} shows the asteroseismic observables used, as well as the \dnu adopted for the GP length scale parameter.
We define GP length scale as an integer multiple of \dnu and, by comparing returned model evidences, we arrive at an optimal value of $7\times\dnu$ ($945.7\muHz$).
\begin{table}
    \centering
    \begin{threeparttable}
    	\caption{Classical observables adopted for the Sun.}
        \label{tab:solar_classicals}        
    	\begin{tabular}{ccc}
    		\hline
    		Parameter & Value & Reference \\
                \hline
                $\Teff$ & $5777 \pm 20$\,K & 1 \\
                $\Lum$ & $1 \pm 0.001\Lodot$ & 2 \\
                $\FeH$ & $0.00 \pm 0.01$\,dex & 3 \\
    		\hline
    	\end{tabular}
        
        \begin{tablenotes}
        \textbf{References:} 1 -- \citet{Scott_2015}, 2 -- \citet{Kopp_2025}, 3 -- \citet{Asplund_2009}
        \end{tablenotes}
        
    \end{threeparttable}
    
\end{table}

\begin{table}
	\centering
    \begin{threeparttable}
    	\caption{Asteroseismic observables adopted for the Sun.}
        \label{tab:solar_asteroseismic}        
    	\begin{tabular}{ccc}
    		\hline
    		Parameter & Freq. [\muHz] & Reference \\
                \hline
                $\dnu$ & $135.1 \pm 0.2$ & 1 \\
                $\nu_{n=6}$ & $972.615 \pm 0.002$ & 2 \\
                $\nu_{n=7}$ & $1117.993 \pm 0.004$ & 2 \\
                $\nu_{n=8}$ & $1263.198 \pm 0.005$ & 2 \\
                $\nu_{n=9}$ & $1407.472 \pm 0.006$ & 2 \\
                $\nu_{n=10}$ & $1548.336 \pm 0.007$ & 2 \\
                $\nu_{n=11}$ & $1686.594 \pm 0.012$ & 2 \\
                $\nu_{n=12}$ & $1822.202 \pm 0.012$ & 2 \\
                $\nu_{n=13}$ & $1957.452 \pm 0.012$ & 2 \\
                $\nu_{n=14}$ & $2093.518 \pm 0.013$ & 3 \\
                $\nu_{n=15}$ & $2228.749 \pm 0.014$ & 3 \\
                $\nu_{n=16}$ & $2362.788 \pm 0.016$ & 3 \\
                $\nu_{n=17}$ & $2496.180 \pm 0.017$ & 3 \\
                $\nu_{n=18}$ & $2629.668 \pm 0.015$ & 3 \\
                $\nu_{n=19}$ & $2764.142 \pm 0.015$ & 3 \\
                $\nu_{n=20}$ & $2899.022 \pm 0.013$ & 3 \\
                $\nu_{n=21}$ & $3033.754 \pm 0.014$ & 3 \\
                $\nu_{n=22}$ & $3168.618 \pm 0.017$ & 3 \\
                $\nu_{n=23}$ & $3303.520 \pm 0.021$ & 3 \\
                $\nu_{n=24}$ & $3438.992 \pm 0.030$ & 3 \\
                $\nu_{n=25}$ & $3574.893 \pm 0.048$ & 3 \\
                $\nu_{n=26}$ & $3710.717 \pm 0.088$ & 3 \\
                $\nu_{n=27}$ & $3846.993 \pm 0.177$ & 3 \\
                $\nu_{n=28}$ & $3984.214 \pm 0.323$ & 3 \\
    		\hline
    	\end{tabular}

        \begin{tablenotes}
        \textbf{References:} 1 -- \citet{Huber_2011}, 2 -- \citet{Hale_2016, Davies_2014}, 3 -- \citet{Hale_2016, Broomhall_2009}
        \end{tablenotes}
        
    \end{threeparttable}

\end{table}

\begin{figure*}
    \centering
    \includegraphics[width=0.6\linewidth]{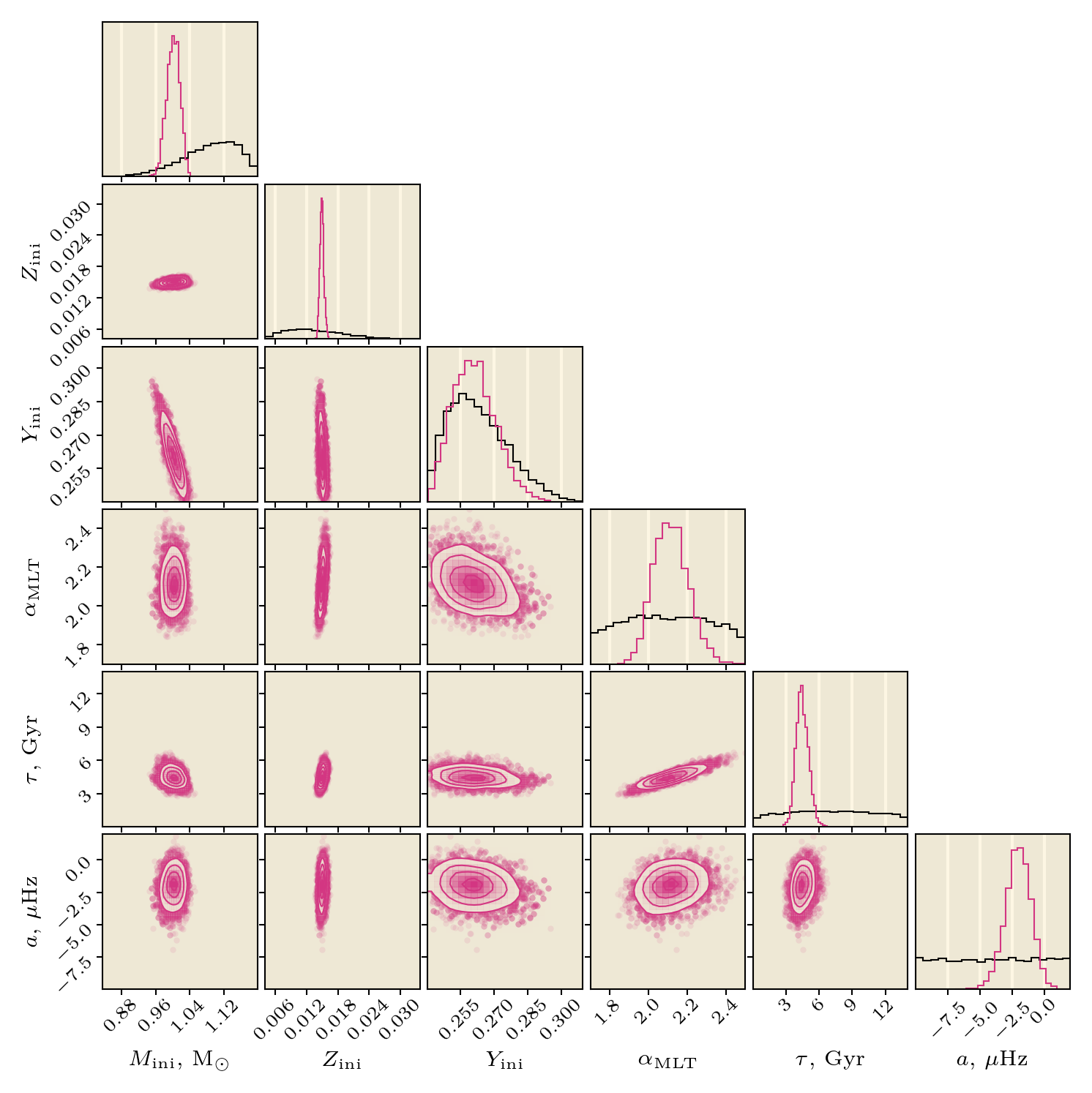}
    \caption{Posterior samples for the Sun shown in pink. The off-diagonal panels show the joint distributions, marginal distributions are shown in the diagonal panels, and samples from the prior distribution are shown in black.}
    \label{fig:posterior_sun}
\end{figure*}

The returned posterior samples are shown in Figure~\ref{fig:posterior_sun}.
From these returned posterior samples, our inferred solar fundamental properties and surface term \KBa parameter are listed in Table \ref{tab:solar_inferred}.
We report a solar mass of $1.00 \pm 0.02\Modot$.
Despite only including \Teff, \Lum, \FeH, and a set of individual radial modes, we have demonstrated our ability to constrain the \Mass parameter to an uncertainty of 2 per cent.
This is despite our rigorous treatment of the random uncertainties, and of the systematics beyond those inherent in the grid model physics assumptions.

\begin{table}
	\centering
	\caption{Returned fundamental and surface correction parameters for the Sun.}
	\begin{tabular}{ccc}
		\hline
		Parameter & Value \\
            \hline
            $\Mass$ & $1.00 \pm 0.02\Modot$ \\
            $\Zini$ & $0.0150 \pm 0.0004$ \\
            $\Yini$ & $0.26 \pm 0.01$ \\
            $\MLT$ & $2.11 \pm 0.09$ \\
            $\Age$ & $4.48 \pm 0.55\Gyr$ \\
            $\KBa$ & $-2.01 \pm 0.98\muHz$ \\
		\hline
	\end{tabular}
    \label{tab:solar_inferred}
\end{table}

We find a solar initial metal mass fraction of $0.0150 \pm 0.0004$.
This is in reasonably good agreement with the value of $0.0142$ found by \citet{Asplund_2009}, which was used in calibrating the grid of stellar models.
We find an even better agreement with the updated value of $0.0154$ from \citet{Asplund_2021}, and fall comfortably within the range of values spanning $0.0130 - 0.0188$ from other compilations of solar chemical compositions \citep[see ][]{Grevesse_1998, Asplund_2005, Lodders_2020}.

Our determination of solar age of $4.48 \pm 0.55\Gyr$ from modelling with individual radial modes is consistent with the helioseismic solar age of $4.57 \pm 0.11\Gyr$ determined by \citet{Bonanno_2002}, and also agrees with the published meteoric solar age of $4.6 \pm 0.1\Gyr$ from \citet{Connelly_2012}.

While \Pitchfork precision far exceeds the observational precision for solar \Teff and \FeH, emulation uncertainty is the dominating factor for \Lum and the individual mode frequencies.
This suggests that improvement on these solar fundamental property constraints is feasible should emulation uncertainty be reduced further.
We aim to address this in future work by using ensemble methods to improve \Pitchfork precision by up to an order of magnitude.

We can also use \Pitchfork to show posterior predictions on the observables parameters in a posterior predictive check.
To do this, we use the returned posterior samples of the fundamental parameters as inputs to our emulator, and show the corresponding predictions on the observables.
For the classical observables, all of which are supplied as observed results during inference, this purely serves as a diagnostic check; if the posterior predicted distributions deviated significantly from the observed values used as inference inputs, this would indicate an error in our sampling and diminish confidence in our posterior.

\begin{figure*}
\begin{subfigure}{.5\textwidth}
  \centering
  \includegraphics[width=0.95\linewidth]{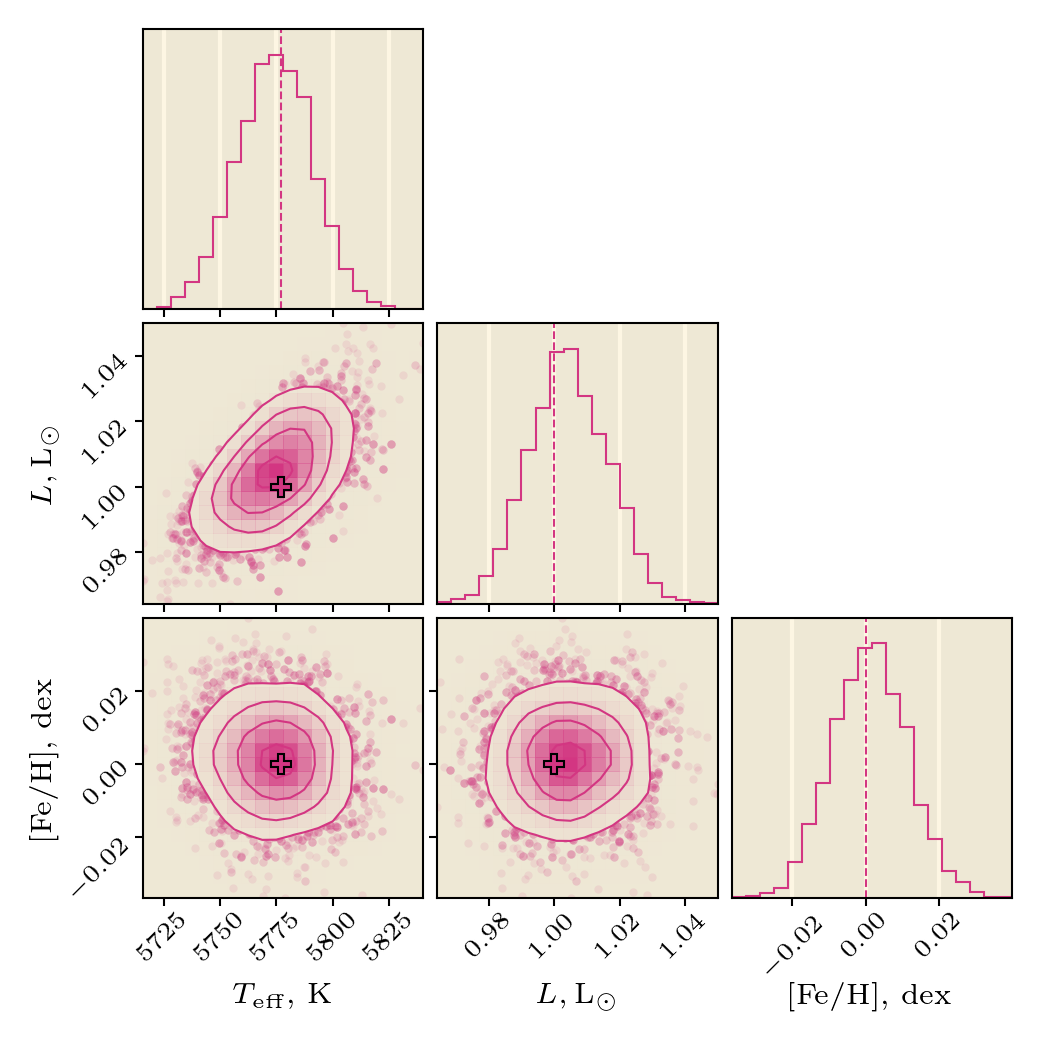}
  \subcaption{Solar posterior predicted classical observables}
  \label{subfig:posterior_predictive_corner_sun}
\end{subfigure}%
\begin{subfigure}{.5\textwidth}
  \centering
  \includegraphics[width=\linewidth]{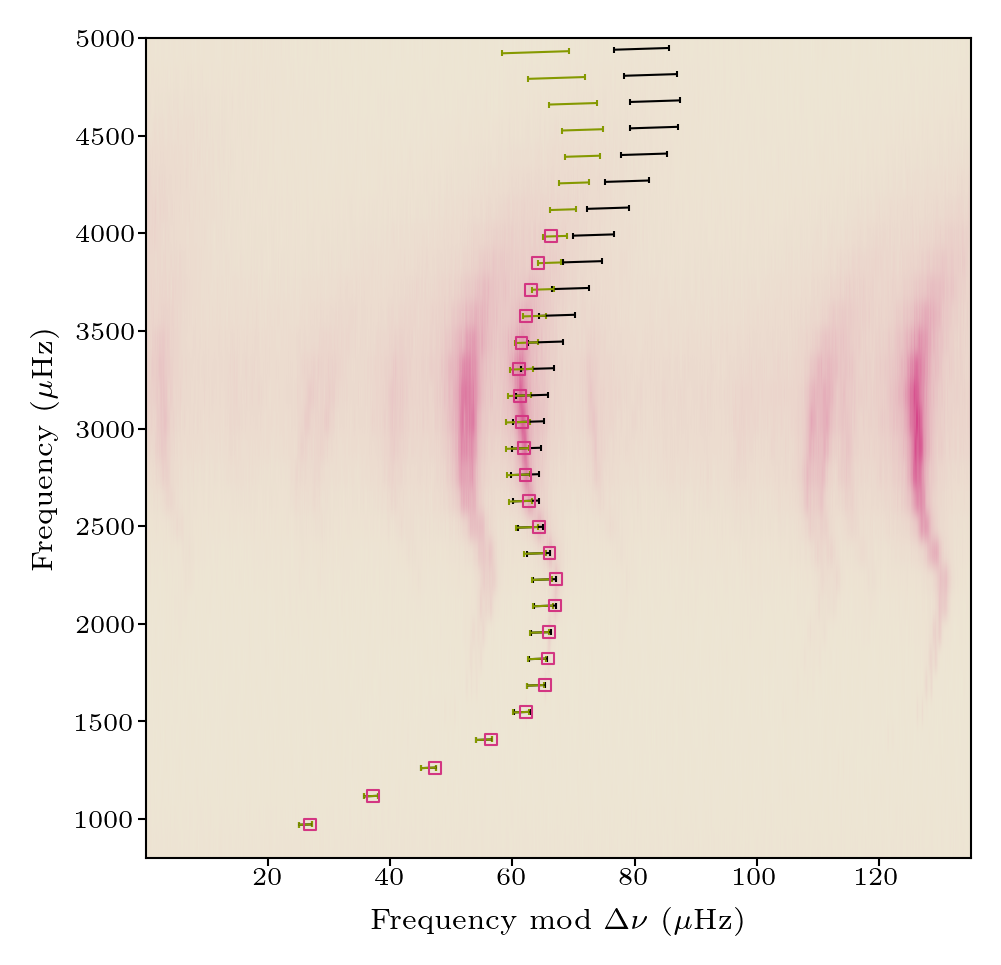}
  \subcaption{Solar posterior predicted frequency échelle diagram}
  \label{subfig:posterior_predictive_echelle_sun}
\end{subfigure}
\caption{Results from the posterior predictive test for the Sun. \textbf{(a)}: Posterior predictive distributions on the classical observables for the Sun, shown in pink. The crosses and dotted lines are the observed values used as inputs for the inference pipeline. \textbf{(b)}: Posterior predicted frequency échelle diagram for the Sun. The solar amplitude spectrum is shown in pink in the background, and identified modes used as inputs for the inference pipeline are shown as pink squares. The black bars show the one sigma range of the posterior predicted frequency distributions without a surface correction, and the green bars show the results of applying surface correction corresponding to the posterior \KBa and \KBb samples.}
\label{fig:posterior_predictive_sun}
\end{figure*}

Figure \ref{subfig:posterior_predictive_corner_sun} shows the posterior predictive distributions on the classical observables from using the posterior samples shown in Figure \ref{fig:posterior_sun}.
We report posterior predicted solar effective temperature of $5775\pm17$\,K, luminosity of $1.00\pm0.01\Lodot$, and surface metallicity of $0.00\pm0.01$\,dex.

Our ability to emulate a full set of radial mode frequencies of orders $6\leq n \leq 40$ for a set of inputs allows us to compare posterior predicted mode frequencies to the full power spectrum.
Figure \ref{subfig:posterior_predictive_echelle_sun} shows the solar échelle spectrum over-plotted with the identified radial modes used in sampling.
For each fundamental parameter posterior sample, we predict every emulated radial mode frequency that the emulator was trained to predict, and show the resulting posterior predictive (shown in black in Figure \ref{subfig:posterior_predictive_echelle_sun}).
Furthermore, our ability to sample the \KBa and \KBb parameters of the surface correction means that each posterior sample has a corresponding surface correction, which can be applied to the posterior predicted mode frequencies (shown in green in Figure \ref{subfig:posterior_predictive_echelle_sun}).
The result is a set of corrected posterior predicted radial mode frequencies of orders $6 \leq n \leq 40$ that agree with the observed modes within the 1$\sigma$ range of the posterior predicted distributions.

\subsubsection{16 Cygni A and B}
\label{sec:16cyg}
Here we present results for the asteroseismic binary system 16 Cygni A and B.
The close proximity, abundance of high-quality \textit{Kepler} data, and presence of the exoplanet 16 Cygni Bb \citep{Cochran_1997} has made this system a popular subject of asteroseismic study \citep[see e.g. ][]{Davies_2015, Bellinger_2017, Lund_2017, Nsamba_2022}.
Additionally, the similarity in mass and compositions of both members of this binary to the Sun makes this a promising benchmark system for understanding systematics in models of stellar evolution.

Despite this being a known binary system, we treated each component entirely independently in order to test our ability to retrieve the expected agreement in returned \Zini and \Age parameters.
Table \ref{tab:16cyg_classicals} shows the adopted classical observables, and Table \ref{tab:16cyg_asteroseismic} shows the asteroseismic observables used, including the \dnu used in the GP kernel length scale definition for the A and B components.
We found optimal GP length scales of $6\times\dnu$ ($619.8$\muHz) and $5\times\dnu$ ($584.5$\muHz) for 16 Cygni A and B, respectively.

\begin{table}
\centering
    \begin{threeparttable}
    	\caption{Classical observables adopted for 16 Cygni A and B.}
        \label{tab:16cyg_classicals}
    	\begin{tabular}{cccc}
    		\hline
    		Parameter & A Value & B Value & Reference \\
                \hline
                $\Teff$ &  $5839 \pm 42$\,K & $5809 \pm 39$\,K & 1 \\
                $\Lum$ & $1.56 \pm 0.05\,L_\odot$ & $1.27 \pm 0.02\Lodot$ & 2 \\
                $\FeH$ & $0.96 \pm 0.026$\,dex & $0.052 \pm 0.021$\,dex & 3 \\
    		\hline
    	\end{tabular}

        \begin{tablenotes}
        \textbf{References:} 1 -- \citet{White_2013}, 2 -- \citet{Metcalfe_2012}, 3 -- \citet{Ramirez_2009}.
        \end{tablenotes}
        
    \end{threeparttable}

\end{table}

\begin{table}
\centering
    \begin{threeparttable}
	\caption{Asteroseismic observables adopted for 16 Cygni A and B.}
    \label{tab:16cyg_asteroseismic}
    	\begin{tabular}{cccc}
    		\hline
    		Parameter & A Freq. [\muHz] & B Freq. [\muHz] & Reference \\
                \hline
                $\dnu$ & $103.3 \pm 0.021$ & $116.9 \pm 0.013$ & 1 \\
                $\nu_{n=12}$ & $1390.808 \pm 0.757$ & \textemdash & 1 \\
                $\nu_{n=13}$ & $1495.053 \pm 0.243$ & $1695.023 \pm 0.141$ & 1 \\
                $\nu_{n=14}$ & $1598.690 \pm 0.075$ & $1812.445 \pm 0.147$ & 1 \\
                $\nu_{n=15}$ & $1700.952 \pm 0.102$ & $1928.886 \pm 0.110$ & 1 \\
                $\nu_{n=16}$ & $1802.351 \pm 0.084$ & $2044.357 \pm 0.071$ & 1 \\
                $\nu_{n=17}$ & $1904.521 \pm 0.059$ & $2159.503 \pm 0.057$ & 1 \\
                $\nu_{n=18}$ & $2007.538 \pm 0.042$ & $2275.949 \pm 0.049$ & 1 \\
                $\nu_{n=19}$ & $2110.950 \pm 0.037$ & $2392.645 \pm 0.046$ & 1 \\
                $\nu_{n=20}$ & $2214.225 \pm 0.055$ & $2509.678 \pm 0.043$ & 1 \\
                $\nu_{n=21}$ & $2317.282 \pm 0.055$ & $2626.458 \pm 0.052$ & 1 \\
                $\nu_{n=22}$ & $2420.937 \pm 0.082$ & $2743.322 \pm 0.066$ & 1 \\
                $\nu_{n=23}$ & $2524.950 \pm 0.148$ & $2860.680 \pm 0.094$ & 1 \\
                $\nu_{n=24}$ & $2628.930 \pm 0.257$ & $2978.180 \pm 0.171$ & 1 \\
                $\nu_{n=25}$ & $2733.571 \pm 0.445$ & $3097.170 \pm 0.414$ & 1 \\
                $\nu_{n=26}$ & $2840.148 \pm 1.058$ & $3216.451 \pm 0.453$ & 1 \\
                $\nu_{n=27}$ & $2944.937 \pm 0.896$ & $3336.009 \pm 1.038$ & 1 \\
    		\hline
    	\end{tabular}

        \begin{tablenotes}
        \textbf{References:} 1 -- \citet{Lund_2017}.
        \end{tablenotes}
        
    \end{threeparttable}

\end{table}

\begin{figure*}
    \centering
    \includegraphics[width=0.6\linewidth]{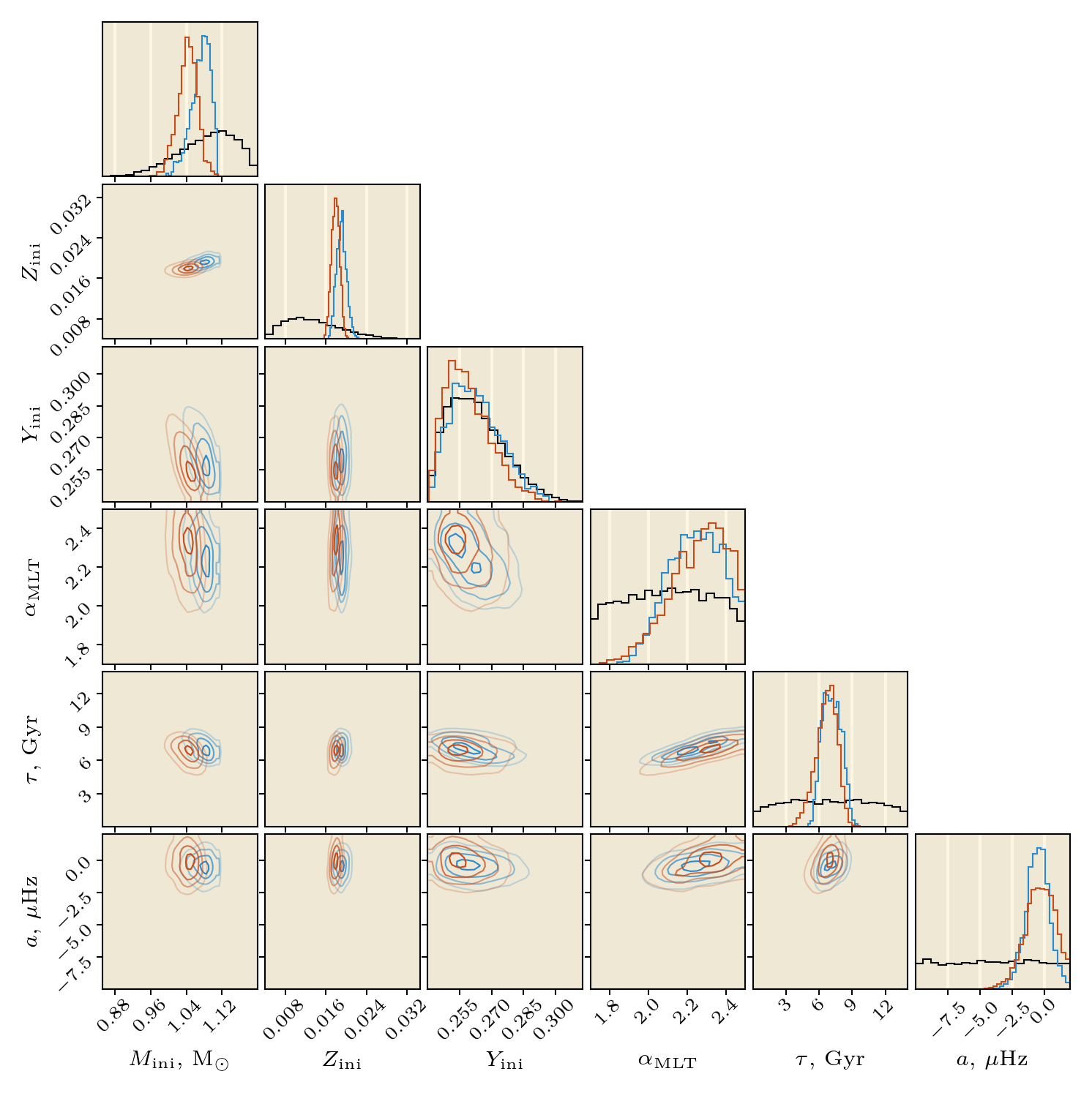}
    \caption{Posterior samples for 16 Cygni A (blue) and B (orange). The off-diagonal panels show the joint distributions, marginal distributions are shown in the diagonal panels, and samples from the prior distribution are shown in black.}
    \label{fig:posterior_16cyg}
\end{figure*}

Figure \ref{fig:posterior_16cyg} shows the over-plotted posterior samples for both A and B, and the inferred fundamental parameters and surface term \KBa parameter is shown in Table \ref{tab:16cyg_inferred}.
Our returned posteriors show agreement in both \Age and \Zini, which is expected for a binary system modelled independently.
This agreement indicates that we could see improved constraints by treating the binary hierarchically, which is an extension of this method that we intend to explore in future work.

Our inferred \Age values for A and B are $7.13\pm0.89\Gyr$ and $6.75\pm0.94\Gyr$, respectively.
These are in agreement with the span of values from the different pipelines in the LEGACY sample \citep[][referred to henceforth as SA17]{Aguirre_2017} of $6.67\text{--}7.52$\,Gyr and $6.92\text{--}7.39$\,Gyr, respectively.
Additionally, we return find a \Mass for A of $1.08\pm0.02\Modot$, which matches well with the results from SA17, which range from $1.05\text{--}1.11\Modot$.
For B, however, we note the discrepancy between our inferred \Mass of $1.04\pm0.02\Modot$ and the LEGACY span of $0.99\text{--}1.02\Modot$.
This could potentially be explained by the differences in the model grids used.
For example, we considered a variable \MLT and \Yini, whereas the \textsc{BASTA} results in SA17 used a fixed solar-calibrated value of \MLT and a linear Galactic enrichment law linking \Yini to \Zini with a fixed slope.
Furthermore, all of the pipelines included in SA17 are based on a higher metallicity solar mixture (either that of \citet{Grevesse_1993} or \citet{Grevesse_1998}), than the \citet{Asplund_2009} model used for the \MESA grid on which \Pitchfork was trained.

We note that the difference in systematic assumptions and methodological approaches makes a like-for-like comparison challenging: it is understood that different systematic assumptions and modelling approaches can influence inferred stellar fundamental properties \citep{Valle_2015, Nsamba_2018}.
As we find here for the results for 16 Cygni A and B, the impact of untreated systematics can influence results to measurably different degrees even for stars that occupy proximate regions of fundamental parameter space.
This highlights an important point: in order to understand how systematics are influencing our inference of stellar fundamental properties, we must first be confident that the sources of random uncertainty are being accounted for correctly.
The ideal method would be capable of scaling to high dimensions, flexible to operating on different grids from different modelling codes, and allow constraint from individual mode frequency measurements, all while being computationally tractable.
\Pitchfork and the inference pipeline described here is an example of such a method, but this deserves dedicated study which we leave to future work.

\begin{table}
	\centering
	\caption{Returned fundamental and surface correction parameters for 16 Cygni A and B.}
	\begin{tabular}{ccc}
		\hline
		Parameter & A Value & B Value \\
            \hline
            $\Mass$ & $1.08 \pm 0.02\,M_\odot$ & $1.04 \pm 0.02\,M_\odot$ \\
            $\Zini$ & $0.019 \pm 0.001$ & $0.018 \pm 0.001$\\
            $\Yini$ & $0.26 \pm 0.01$ & $0.26 \pm 0.01$\\
            $\MLT$ & $2.25 \pm 0.15$ & $2.28 \pm 0.15$\\
            $\Age$ & $7.13 \pm 0.89$\,Gyr & $6.75 \pm 0.94$\,Gyr\\
            $\KBa$ & $-0.48 \pm 0.86\,\mu$Hz & $-0.26 \pm 1.15\,\mu$Hz\\
		\hline
	\end{tabular}
    \label{tab:16cyg_inferred}
\end{table}

\begin{figure*}
\begin{subfigure}{.5\textwidth}
  \centering
  \includegraphics[width=\linewidth]{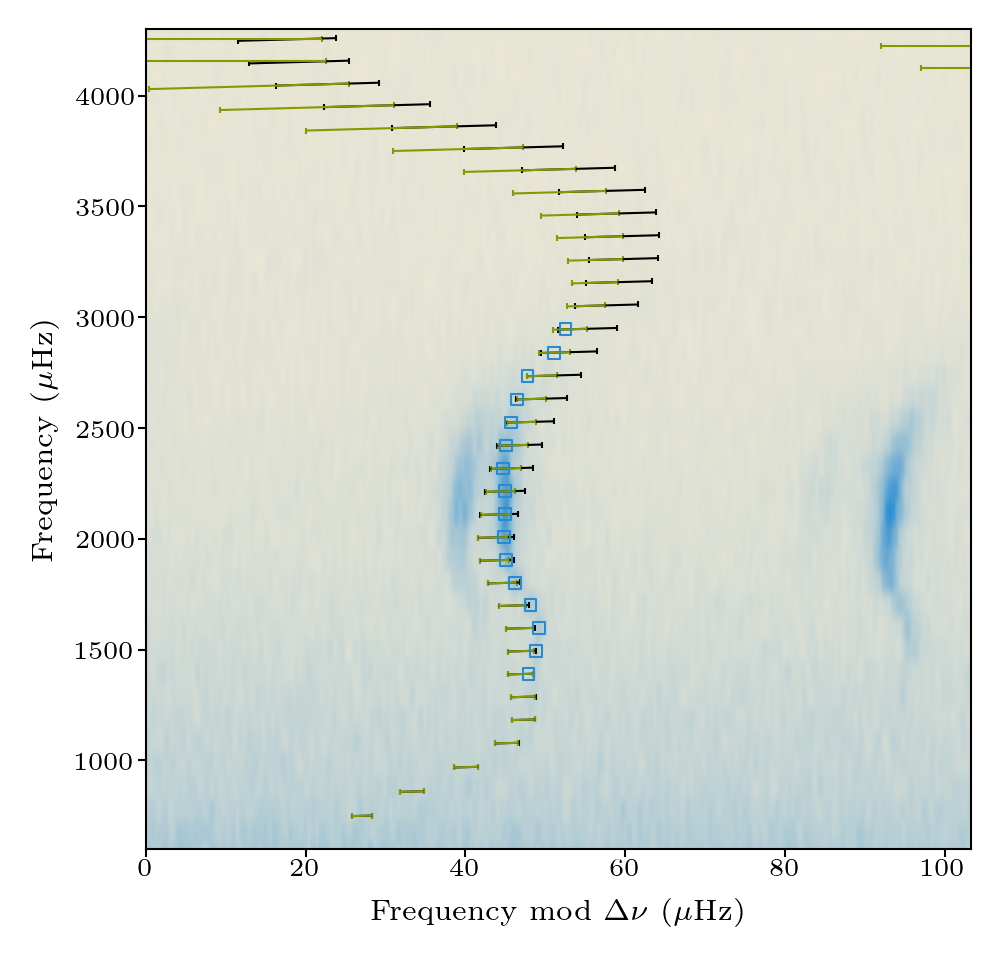}
  \subcaption{16 Cygni A posterior predicted frequency échelle diagram}
  \label{subfig:posterior_predictive_echelle_16CygA}
\end{subfigure}%
\begin{subfigure}{.5\textwidth}
  \centering
  \includegraphics[width=\linewidth]{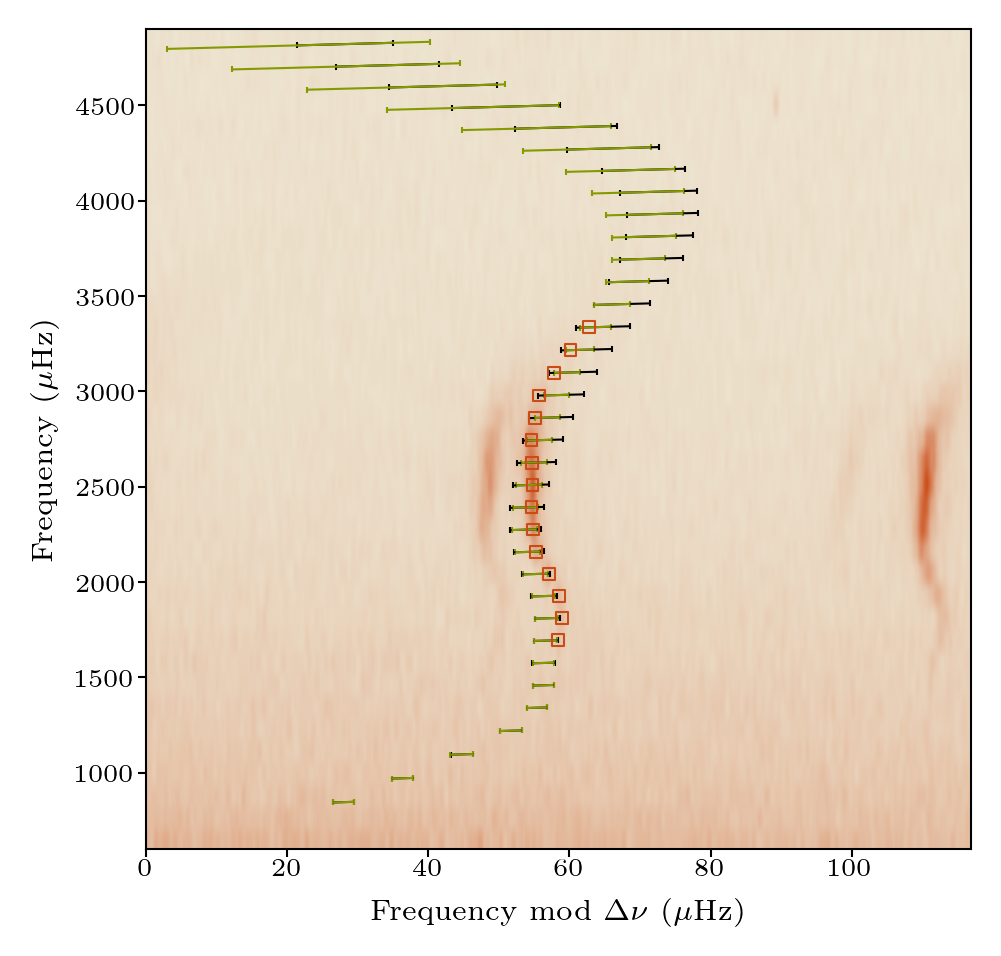}
  \subcaption{16 Cygni B posterior predicted frequency échelle diagram}
  \label{subfig:posterior_predictive_echelle_16CygB}
\end{subfigure}
\caption{Posterior predicted frequency échelle diagrams for the 16 Cygni A (left, blue) and B (right, orange). The amplitude spectrums are shown in the background, and identified modes used as inputs for the inference pipeline are are shown as coloured squares. The black bars show the one sigma range of the posterior predicted frequency distributions without a surface correction, and the green bars show the results of the applying surface correction corresponding to the posterior \KBa and \KBb samples.}
\label{fig:posterior_predictive_16Cyg}
\end{figure*}

Figure \ref{fig:posterior_predictive_16Cyg} shows the posterior predicted frequencies for both 16 Cygni A and B compared to the power spectrum and the identified modes used as inputs for sampling.
The \KBa and \KBb samples do correct the emulated posterior predicted frequencies to some degree, but there is still disagreement between corrected posterior predicted and observed frequencies for both 16 Cygni A and B.
This indicates that a more complex treatment of the surface term could improve inference -- for example, the prescription described by \citet{Ball_2014}.
This would require training a new emulator that is capable of predictions of the mode inertias as well as the individual modes, which we leave to future work.
We also highlight the presence of unaddressed systematic differences between emulated radial modes and observed modes that are not due to the surface effect alone; the residuals between the uncorrected and the observed frequencies for both 16 Cygni A and B do not increase in magnitude with increasing frequency and consistent \KBa sign, as we'd expect if this were the case.
Rather, \KBa appears to change sign close to $\numax$.
It is not yet clear whether this behaviour arises from untreated systematics in the underlying model grid -- subsequently learned and propagated by \Pitchfork -- or from limitations in our present treatment of the surface term.
To distinguish between these possibilities, future work could involve training emulators on alternative grids, enabling Bayesian model-evidence comparisons in the first case, and developing an emulator capable of predicting non-radial modes or inertiae in the second, thereby permitting a more comprehensive treatment of the surface term.

\section{Conclusions}
\label{sec:conclusions}
In this paper, we have presented a method for training a neural network as an emulator of the \textsc{MESA} stellar modelling and \textsc{GYRE} stellar oscillation codes.
\Pitchfork, our neural network emulator with a branching architecture, is capable of emulating individual radial mode frequencies as a rapid and trivially scalable alternative to interpolation or on the fly modelling.
We have shown how \Pitchfork can be used for vectorised likelihood evaluation during nested sampling of stellar fundamental parameter posterior distributions.
We have tested our method in an extensive hare-and-hounds exercise, and have shown examples of recovered posterior samples for benchmark stars -- namely, the Sun and the asteroseismic binary 16 Cygni A and B.
For these three stars, we have demonstrated the ability to retrieve fundamental stellar parameter posterior samples that match well with published values, and have compared the posterior predictive frequencies with the observed oscillation frequency spectra.

The main conclusions of the paper are as follows:
\begin{itemize}
    
    \item \Pitchfork is capable of emulating individual radial modes of orders $6 \leq n \leq 40$ of solar-like oscillators with masses below $1.2\Modot$ to a consistent percentage uncertainty $0.02$ per cent ($\sigma_{n=6,\,\psi} = 0.3\muHz$, $\sigma_{n=40,\,\psi} = 1.1\muHz$) -- emulation of individual modes of oscillation for solar-like oscillators to this precision is a novel result in itself.
    \item \Pitchfork can predict the classical observables to average precisions of $\TeffNNError = 5.9$\,K, $\LumNNError = 0.014\,\Lodot$, $\FeHNNError = 0.00065\dex$.
    \item Despite the flexibility to include 35 observed individual modes of oscillation in the modelling pipeline, this method does not come at a heavy computational cost due to favourable scaling of neural networks towards vectorisation. \Pitchfork prediction times are on the order of 10~ms for a single point, but only 900~ms for one million points.
    \item We used the emulator for vectorised evaluation of a multivariate Gaussian likelihood function in the \Ultranest nested sampling code -- the result is a statistically rigorous, rapid inference pipeline capable of returning constraints on the stellar fundamental and surface correction parameters and Bayesian model evidences, typically in $60-600$ seconds.
    \item We employed a Gaussian process for treatment of the surface correction, which defines a flexible probability distribution over the functional form of the deviation between modelled and observed frequencies.
    \item We have discussed the anticipated improvements and extensions to this method, including improved precision and point-by-point uncertainty estimation using ensemble approaches, emulation of non-radial modes of oscillation, and the use of Bayesian model evidences to characterise systematics.
    \item From an extensive hare-and-hounds exercise, we have demonstrated that high-sigma draws of observational noise will correctly influence returned posterior samples and, on a population scale, our inferred values are consistent with the truth values.
    \item We returned solar fundamental parameter values of $\Mass = 1.00 \pm 0.02\Modot$, $\Zini = 0.0150 \pm 0.0004$, and $\Age = 4.48 \pm 0.55\Gyr$, and constrain the surface correction $\KBa$ coefficient to $\KBa = -2.01 \pm 0.98\muHz$.
    \item For the 16 Cygni system, we reported inferred masses of $1.08 \pm 0.02\,M_\odot$ and $1.04 \pm 0.02\,M_\odot$ for the A and B component, respectively, which are in good agreement with published values. Furthermore, we are able to reproduce the expected agreement in \Zini and \Age posteriors for the two binary components, despite independent treatment.
\end{itemize}

Proper treatment of the sources of error inherent in stellar modelling is vital to be able to address the systematic uncertainty arising from imperfect model physics assumptions used in generating models of stellar evolution.
With the exception of our inability to evaluate emulator uncertainty on a point-by-point basis, which we aim to rectify in future work using ensemble methods, we have demonstrated a statistically sound treatment of random uncertainty throughout this work.
Therefore, we believe that this work is a significant step forwards in utilising asteroseismic data to constrain stellar fundamental properties, and paves the way for proper treatment of systematics, which is extremely important in preparation for the abundance of asteroseismic data expected from future missions.

\section*{Acknowledgements}
Firstly, we would like to thank the referee and the scientific editor for their insightful comments which have markedly improved this paper.
OJS, GRD, and AJL acknowledge the support of the Science and Technology Facilities Council.
GRD, AS, EJH, and MBN acknowledge support from the European Research Council (ERC) under the European Union’s Horizon 2020 research and innovation programme (CartographY; grant agreement ID~804752). MBN acknowledges support from the UK Space Agency.
TL acknowledges support from the National Natural Science Foundation of China (NSFC) grant 12373031.
MNL acknowledges support from the ESA PRODEX Programme.
TRB acknowledges support from the Australian Research Council (Laureate Fellowship FL220100117).

\section*{Data Availability}
\Pitchfork and notebooks showcasing examples of inference are available in an MIT licensed public GitHub repository (\url{https://github.com/ojscutt/pitchfork}).
The grid of stellar models used to train \Pitchfork will be supplied upon reasonable request to the authors.

\section*{Software}

Additional software employed in this study, but not explicitly mentioned above, is presented here:
\begin{itemize}
    \item \textsc{Python} \citep{Van_Rossum_1995}
    \item \textsc{matplotlib} \citep{Hunter_2007}
    \item \textsc{NumPy} \citep{Harris_2020}
    \item \textsc{SciPy} \citep{Virtanen_2020}
    \item \textsc{Astropy} \citep{Price_Whelan_2022}
    \item \textsc{Pandas} \citep{Reback_2020}
    \item \textsc{corner} \citep{Foreman_Mackey_2016}
    \item \textsc{echelle} \citep{Hey_2020}
    \item \textsc{scikit-learn} \citep{Pedregosa_2011}
    \item \textsc{Keras} \citep{Chollet_2015}
    \item \textsc{JAX} \citep{Bradbury_2018}
    \item \textsc{tinygp} \citep{Foreman_Mackey_2023}
    \item \textsc{Scientific colour maps} \citep{Crameri_2023}
\end{itemize}



\bibliographystyle{mnras}
\bibliography{main} 



\appendix
\section{Additional Material}
\label{sec:additional_material}
\begin{figure}
    \centering
    \includegraphics[width=\linewidth]{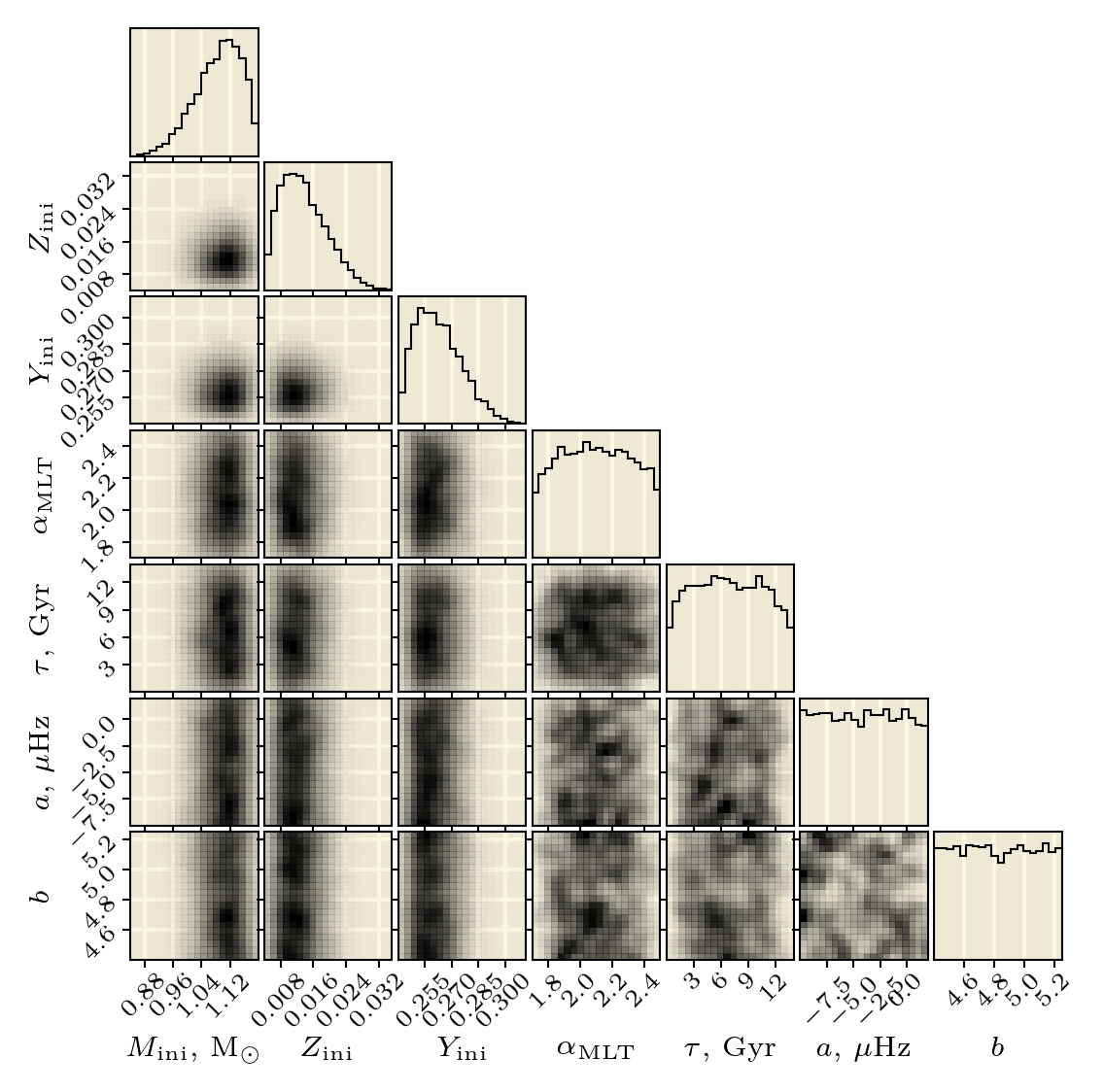}
    \caption{Samples from the prior distribution over model fundamental parameters and surface terms \KBa and \KBb.}
    \label{fig:prior_samples}
\end{figure}

\begin{figure*}
\begin{subfigure}{.5\textwidth}
  \centering
  \includegraphics[width=\linewidth]{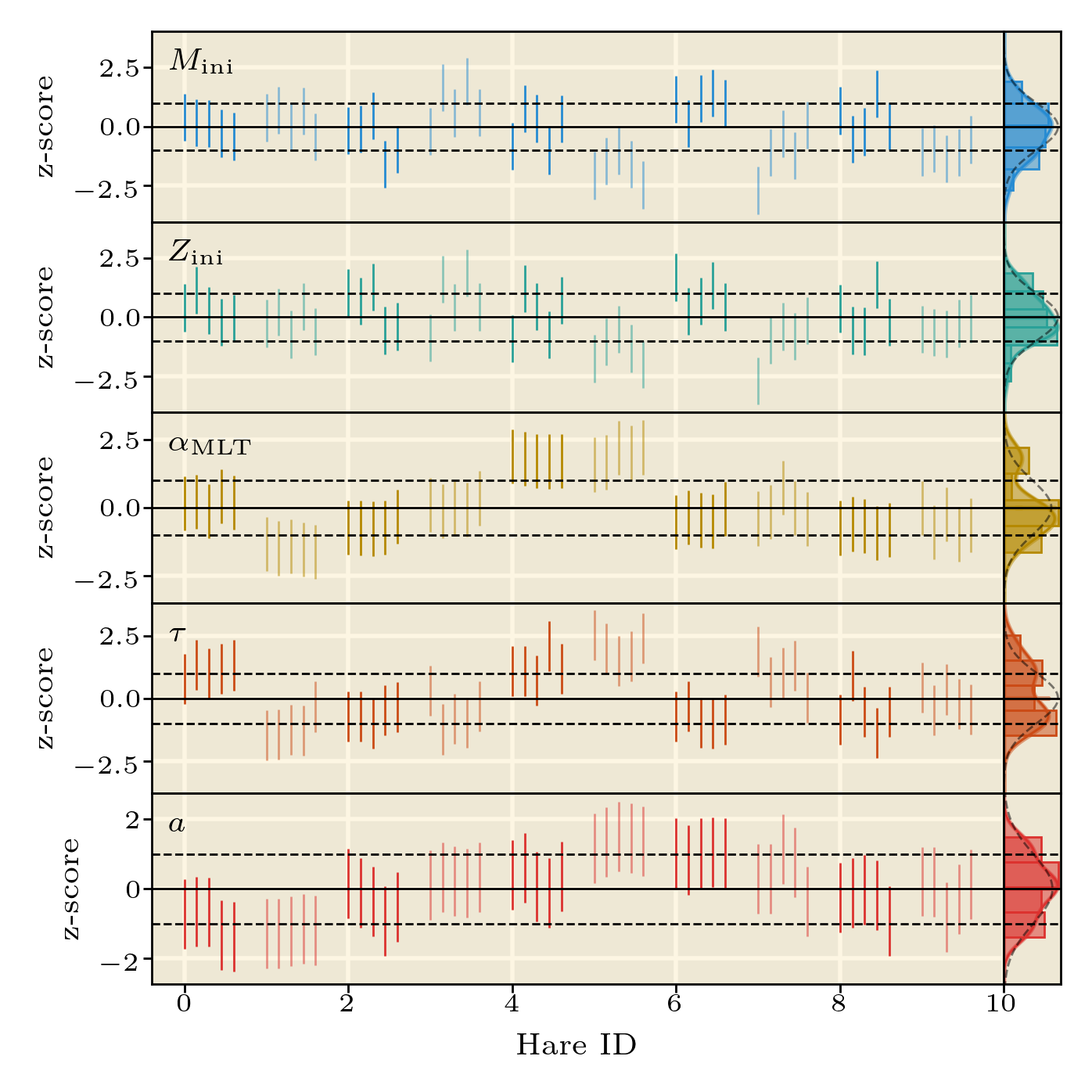}
  \label{subfig:10_scores_hares}
\end{subfigure}%
\begin{subfigure}{.5\textwidth}
  \centering
  \includegraphics[width=\linewidth]{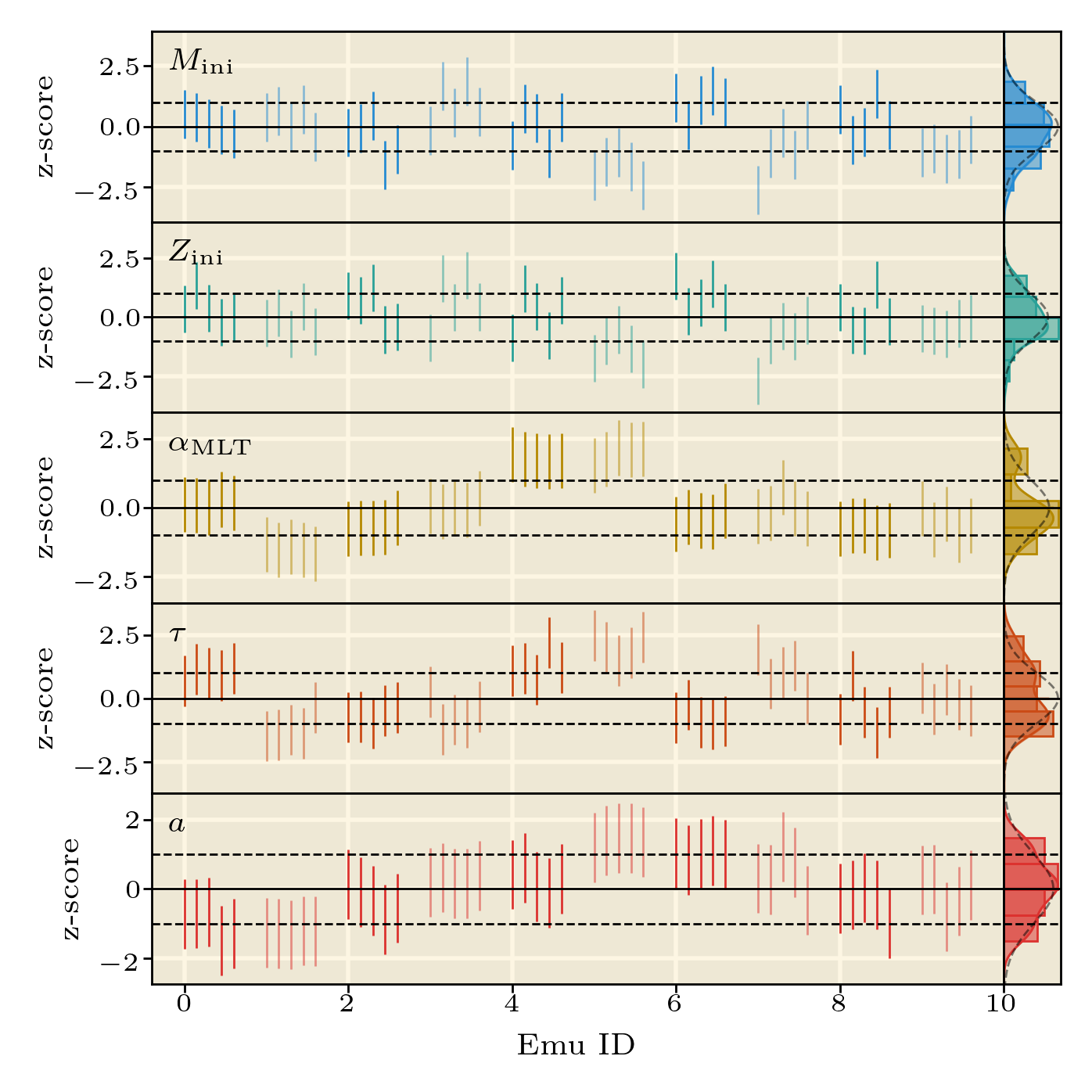}
  \label{subfig:10_scores_emus}
\end{subfigure}
\caption{Posterior $z$-scores over a population of 10 hares (left) and emus (right), each with five draws of observational noise. Each vertical line represents one draw of the observational noise. We alternate the saturation of $z$-score lines for clarity. The histograms and kernel density estimates of all returned $z$-scores are shown on the right hand side, with the target $\mathcal{N}(0,1)$ shown by the dotted grey line.}
\label{fig:10_zscores}
\end{figure*}

In the following, we present additional figures and tables which are referenced in the main text.
This includes a full table of \Pitchfork prediction uncertainties in Table \ref{tab:pitchfork_errors}, samples from the prior distribution on the fundamental properties and surface term parameters in Figure \ref{fig:prior_samples}, and z-score spans for an emu-and-hounds exercise with corresponding hare results for comparison in Figure \ref{fig:10_zscores}.

\begin{table}
	\centering
	\caption{\Pitchfork prediction metrics. $\sigma$ is taken as the standard deviation of the distribution of prediction residuals over the test set. $\sigma_{\%}$ is the mean per cent error of the test set residuals.}
	\begin{tabular}{ccc}
		\hline
		Parameter & $\sigma$ & $\sigma_{\%}$ [per cent] \\
            \hline
            \Teff & $5.893\,\text{K}$ & $0.059$ \\
            \Lum & $0.014\Lodot$ & $0.213$ \\
            \FeH & $0.001\,\text{dex}$ & $0.578$ \\
            $\nu_{n=6}$ & $0.316\muHz$ & $0.035$ \\
            $\nu_{n=7}$ & $0.368\muHz$ & $0.036$ \\
            $\nu_{n=8}$ & $0.381\muHz$ & $0.032$ \\
            $\nu_{n=9}$ & $0.345\muHz$ & $0.027$ \\
            $\nu_{n=10}$ & $0.380\muHz$ & $0.027$ \\
            $\nu_{n=11}$ & $0.360\muHz$ & $0.023$ \\
            $\nu_{n=12}$ & $0.379\muHz$ & $0.023$ \\
            $\nu_{n=13}$ & $0.383\muHz$ & $0.021$ \\
            $\nu_{n=14}$ & $0.409\muHz$ & $0.021$ \\
            $\nu_{n=15}$ & $0.411\muHz$ & $0.020$ \\
            $\nu_{n=16}$ & $0.432\muHz$ & $0.020$ \\
            $\nu_{n=17}$ & $0.441\muHz$ & $0.019$ \\
            $\nu_{n=18}$ & $0.465\muHz$ & $0.019$ \\
            $\nu_{n=19}$ & $0.483\muHz$ & $0.018$ \\
            $\nu_{n=20}$ & $0.489\muHz$ & $0.018$ \\
            $\nu_{n=21}$ & $0.520\muHz$ & $0.018$ \\
            $\nu_{n=22}$ & $0.549\muHz$ & $0.019$ \\
            $\nu_{n=23}$ & $0.565\muHz$ & $0.019$ \\
            $\nu_{n=24}$ & $0.584\muHz$ & $0.019$ \\
            $\nu_{n=25}$ & $0.618\muHz$ & $0.019$ \\
            $\nu_{n=26}$ & $0.657\muHz$ & $0.020$ \\
            $\nu_{n=27}$ & $0.653\muHz$ & $0.019$ \\
            $\nu_{n=28}$ & $0.708\muHz$ & $0.020$ \\
            $\nu_{n=29}$ & $0.720\muHz$ & $0.019$ \\
            $\nu_{n=30}$ & $0.743\muHz$ & $0.019$ \\
            $\nu_{n=31}$ & $0.811\muHz$ & $0.020$ \\
            $\nu_{n=32}$ & $0.802\muHz$ & $0.019$ \\
            $\nu_{n=33}$ & $0.890\muHz$ & $0.020$ \\
            $\nu_{n=34}$ & $0.910\muHz$ & $0.019$ \\
            $\nu_{n=35}$ & $0.930\muHz$ & $0.019$ \\
            $\nu_{n=36}$ & $1.039\muHz$ & $0.020$ \\
            $\nu_{n=37}$ & $0.977\muHz$ & $0.018$ \\
            $\nu_{n=38}$ & $1.070\muHz$ & $0.020$ \\
            $\nu_{n=39}$ & $1.062\muHz$ & $0.019$ \\
            $\nu_{n=40}$ & $1.123\muHz$ & $0.020$ \\
		\hline
	\end{tabular}
 \label{tab:pitchfork_errors}
\end{table}

\section{Justification for the GP}
\label{sec:gp_justification}
\begin{figure*}
\begin{subfigure}{.5\textwidth}
  \centering\includegraphics[width=0.9\linewidth]{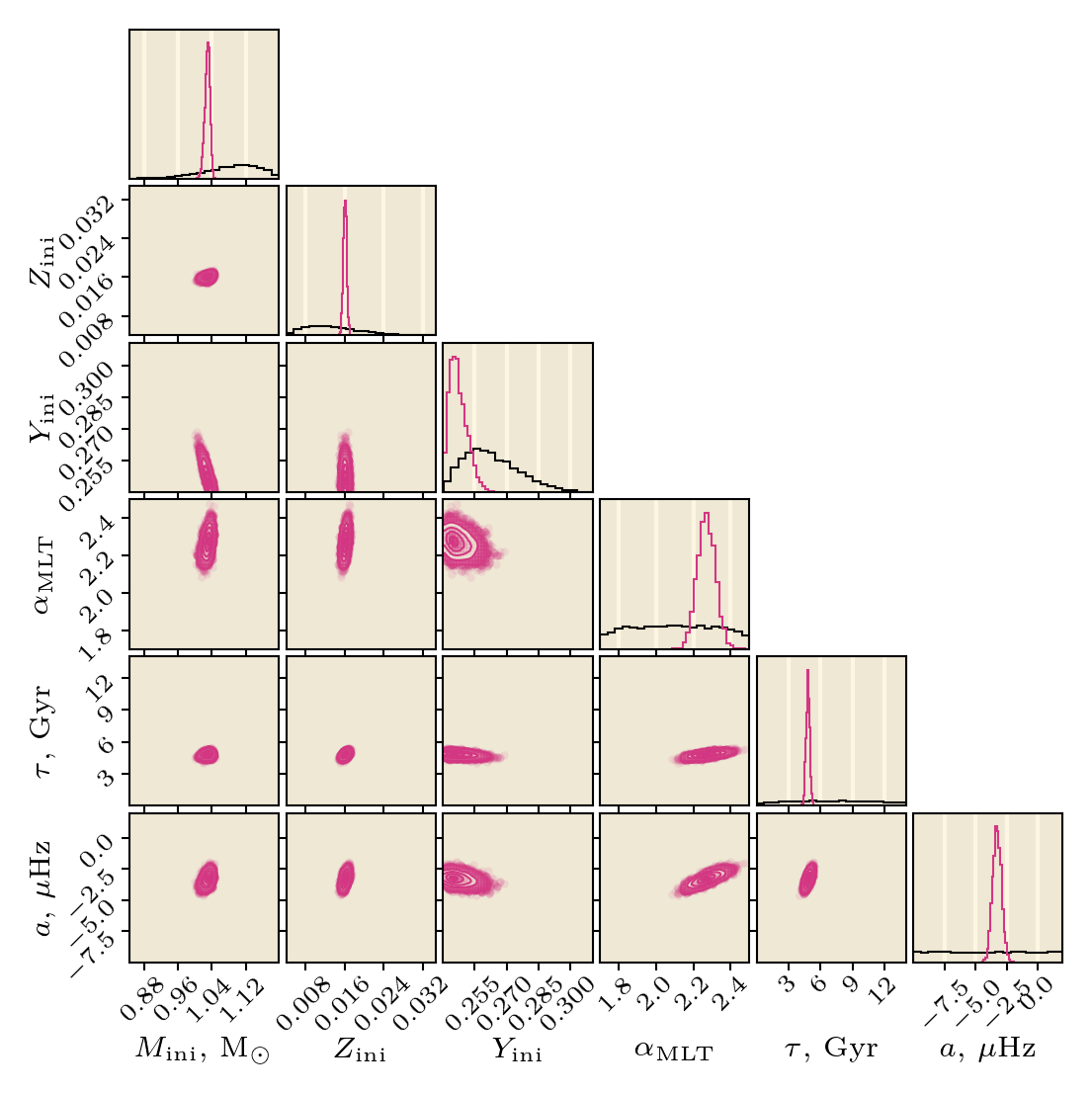}
  \label{subfig:gp_test}
\end{subfigure}%
\begin{subfigure}{.5\textwidth}
    \centering
    \includegraphics[width=0.9\linewidth]{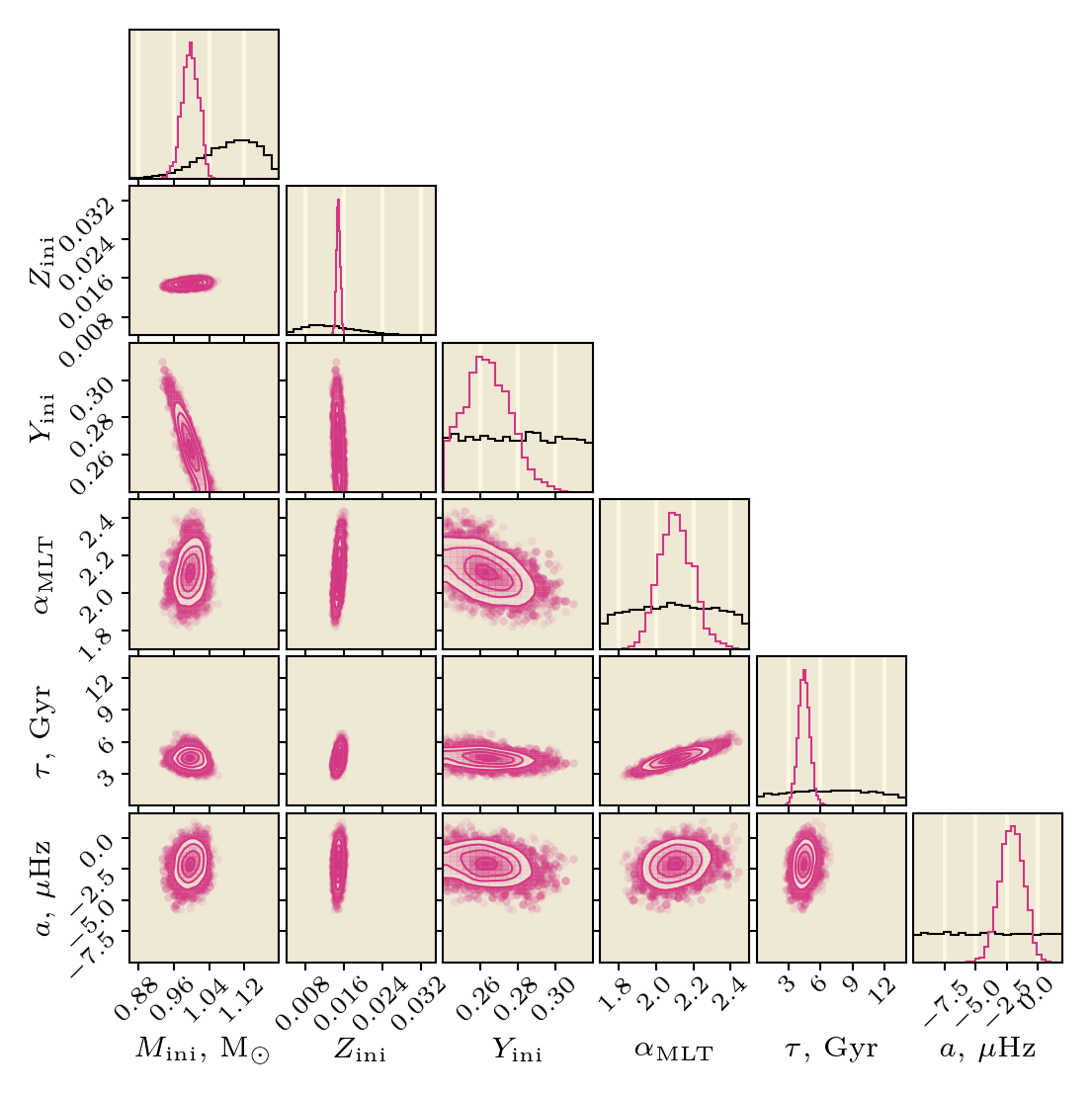}
    \label{subfig:uniform_yini}
\end{subfigure}
\caption{Corner plot showing posterior distribution for solar fundamental properties when GP variance is set to $0\muHz^2$ (left). Corner plot showing posterior distribution for solar fundamental properties when using a uniform \Yini prior (right).}
\label{fig:gp_justification}
\end{figure*}

In this section, we briefly discuss the results of a test conducted to justify the use of the GP approach to modelling the correlated error expected from using an imperfect surface correction.
We performed a Bayesian model evidence comparison between two sampling runs for the Sun.
The first yielded the results presented in Section \ref{sec:sun}.
The second is identical, save for the fact that we neglected the contribution to the random uncertainty budget in the likelihood function from the GP correlated noise model (i.e. by setting the GP variance to zero).
The returned posterior samples for this test can be seen in Figure \ref{fig:gp_justification}.

The reader will notice that the precision on the fundamental property posterior distributions is considerably better than with the correlated noise model.
However, this does not necessarily mean the non-GP model is preferred.
To investigate this, we can compare the returned model log-evidences ($\text{log}Z$) values between the two sampling runs.
Despite better precision on the non-GP model, the evidence is significantly lower ($\text{log}Z = -36$) than the model with the correlated noise model that we presented ($\text{log}Z = -19$). 
This suggests that the model with the GP explains the data better than the model without by a log Bayes factor of 17.
Failing to account for the fact that a parametric surface correction, like the \cite{Kjeldsen_2008} approach used here, is inherently imperfect means we return confidently inaccurate posteriors that do not explain the observed data well.

We also include here the discussion of whether the GP is capable of absorbing glitch signatures, resulting in poor constraint on \Yini (see Section \ref{sec:exemplar_hare}).
Firstly, we consistently use GP length scale values that are far greater than the expected length scale of glitch signatures: the helium ionisation zone glitch varies rapidly as a function of frequency, and would have a much shorter length scale than the values of $>5\dnu$ used in this work.
Additionally, we find that we are able to constrain \Yini somewhat even when using an uninformative uniform prior on \Yini to nearly the same degree as when using the more informative beta used in this work, as can be seen in Figure \ref{fig:gp_justification}.

This suggests that we are able to constrain \Yini using whatever information can be gleaned from the glitch signature present in the radial modes alone.
Including more angular degrees would improve this constraint, and we propose that the combination of a branching neural network architecture and dimensionality re-projection layer on outputs that are highly correlated lends itself very well towards future emulators predicting non-radial modes to this end.
This is an extension we intend to explore in future work.

\bsp	
\label{lastpage}
\end{document}